\documentclass[twocolumn]{aa} 
\usepackage[varg]{txfonts}

\usepackage{natbib}
\usepackage{graphicx}
\usepackage{txfonts}
\usepackage{caption}
\usepackage{multirow}

\usepackage{hyperref}
\def\hii{\mbox{H\,{\sc ii}}}
\def\CF+{\mbox{CF$^+$}}
\def\kms{\mbox{~km\,s$^{-1}$}}
\begin{document}

   \title{Class~I CH$_3$OH Maser Emission from Bar-Driven Inflow Colliding with the Central Molecular Zone}
   \author{V. S. Veena\inst{1,2} \and
    W.-J. Kim\inst{2} \and
    P. Schilke\inst{3} \and
    {\'A}. S{\'a}nchez-Monge\inst{4,5} \and    
    C. Henkel\inst{2} \and
    S. Viti\inst{6,7}\and
    G. Esplugues\inst{8}\and
    F. Wyrowski\inst{2} \and
    W.~E.~Banda-Barrag\'an\inst{9,10} \and
    M. C. Sormani\inst{11} \and          
    D. Riquelme\inst{12} \and
    G. A. Fuller\inst{13}     
    }

    \institute{Department of Astronomy \& Astrophysics, Tata Institute of Fundamental Research,
       Homi Bhabha Road, Mumbai 400005, India \\  \email{veena.s@tifr.res.in}
\and Max Planck Institute for Radioastronomy (MPIfR), Auf dem H\"ugel 69, 53121 Bonn, Germany
\and
    I. Physikalisches Institut, Universit\"at zu Köln, Z\"ulpicher Str. 77, 50937 K\"oln, Germany
    \and
    Institut de Ci\`encies de l'Espai (ICE, CSIC), Carrer de Can Magrans s/n, E-08193, Bellaterra, Barcelona, Spain
    \and
    Institut d'Estudis Espacials de Catalunya (IEEC), Barcelona, Spain
    \and
    Leiden Observatory, Leiden University, PO Box 9513, 2300 RA Leiden, The Netherlands    
    \and
    Transdisciplinary Research Area (TRA) \lq Matter\rq/Argelander-Institut f\"ur Astronomie, University of Bonn
    \and
    Observatorio Astron\'omico Nacional (OAN, IGN), Alfonso XII, 3, 28014 Madrid, Spain   
    \and
    Escuela de Ciencias F\'isicas y Nanotecnolog\'ia, Universidad Yachay Tech, Hacienda San Jos\'e S/N, 100119 Urcuqu\'i, Ecuador
    \and
    Hamburger Sternwarte, University of Hamburg, Gojenbergsweg 112, 21029 Hamburg, Germany
    \and
    Como Lake centre for AstroPhysics (CLAP), DiSAT, Universit{\`a} dell’Insubria, via Valleggio 11, 22100 Como, Italy
    \and
    Departamento de Astronom\'ia, Universidad de La Serena, Av. Cisternas 1200, La Serena, Chile
    \and
    Jodrell Bank Centre for Astrophysics, Department of Physics and Astronomy, The University of Manchester, Manchester M13 9PL, UK    
}


    \date{Received ; accepted }

 
  \abstract
   {The Central Molecular Zone (CMZ) of the Milky Way is shaped by the interplay of bar-driven inflows, shocks, and star formation. At Galactic longitude $l=1.3^\circ$, gas inflowing along the near-side dust lane has been proposed to interact with the CMZ boundary and overshoot into the region above the Galactic plane, making this a key site to investigate how large-scale gas dynamics regulates star formation.}
   {We aim to investigate the presence of Class~I methanol maser emission in this transitional region, testing whether large-scale gas interactions in the CMZ can trigger widespread maser activity via star formation or shocks.}
   {We conducted a dedicated search for the 36.2 and 44.1 GHz Class~I CH$_3$OH maser lines, along with the 48.4 GHz thermal transition, using the Yebes 40m telescope. We complemented these data with archival data from the Herschel-HiGAL survey and the CHIMPS2 survey to explore links between masers, shocks, and star formation.}
   {We detect widespread 36.2 GHz maser emission and two candidate 44.1 GHz masers in a region extending several parsecs. The brightest 36.2 GHz maser has an isotropic luminosity $0.9\times10^{-3}$ L$_\odot$, placing it among the most luminous Galactic Class~I masers. Thermal CH$_3$OH emission and SiO (1--0) emission extend over mapped area of $\sim$24 pc, with both species showing enhanced fractional abundances relative to H$_2$. CO(3--2) position--velocity analysis further shows that the masers are associated with an extended velocity feature (EVF) at $\mathrm{V_{LSR}}\sim$100~\kms.}
   {We conclude that the observed masers are primarily associated with shock-processed gas in a kinematically complex bar--CMZ interface region. Large-scale gas interactions are likely to play an important role in producing the maser emission, although a subset of the masers may also be linked to shocks driven by local star-formation activity. This region therefore provides a promising Galactic analogue of shock-dominated Class~I CH$_3$OH maser environments observed in nuclear regions of barred galaxies.} 


   \maketitle
%

\section{Introduction}

The Galactic Centre (GC) is a complex and dynamic region shaped by the interplay of the supermassive black hole SgrA*, large-scale gas flows, star formation, and feedback \citep{{1983Natur.301..661T},{1991MNRAS.252..210B},{1993ApJ...408..496M},{2005ApJ...620..744G}}. At the heart of this region lies the Central Molecular Zone (CMZ), a dense reservoir of molecular gas spanning $\sim$400~pc, located within the inner kiloparsec of the Milky Way \citep{1996ARA&A..34..645M}. The CMZ harbours a substantial fraction of the Galaxy’s dense molecular gas \citep[][and references therein]{2023ASPC..534...83H} and is thought to be fed by bar-driven inflows that funnel material inward along the Galactic bar dust lanes. On the positive longitude side, one such inflow is associated with the near-side dust lane, which is believed to intersect the CMZ at Galactic longitude $l=1.3^\circ$ (hereafter G1.3). This location has been proposed as the interface where inflowing gas from the dust lane collides with the outer edge of the CMZ \citep[e.g.,][]{2019MNRAS.488.4663S}. Rather than triggering widespread star formation, observations suggest that the turbulent dynamics and shear forces at this junction may suppress star formation \citep{2022A&A...668A.183B}, making the G1.3 region a key site for studying the physical conditions governing star formation regulation in barred spiral galaxies.

\begin{figure*}[!htb]
\centering
\includegraphics[scale=0.35]{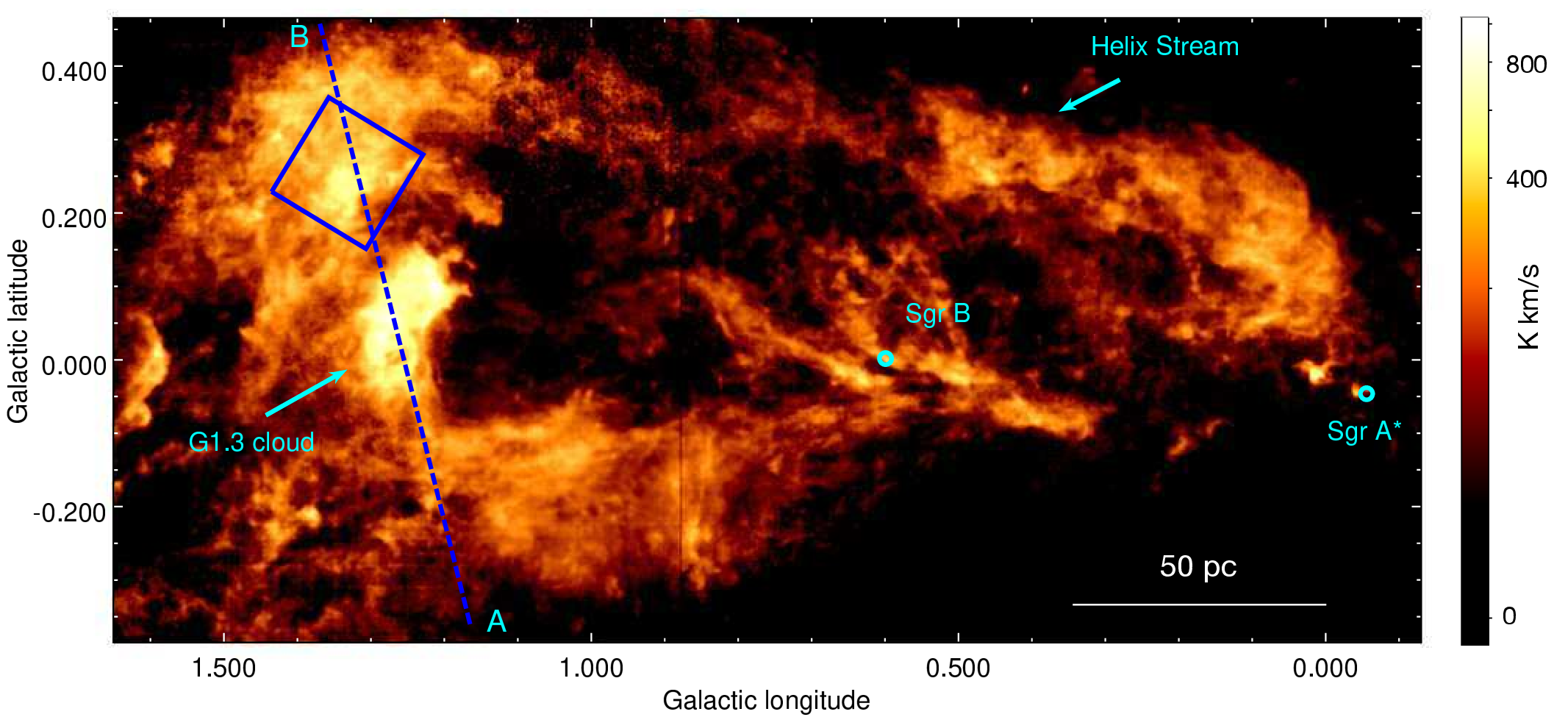}
\caption{Integrated intensity map of $^{12}$CO (3–2) in the Local Standard of Rest (LSR) velocity range 100 to 200\kms~from the CHIMPS2 survey \citep{2020MNRAS.498.5936E}. Locations of the Helix stream and the G1.3 cloud are indicated with arrows. Positions of Sgr A* and peak of Sgr B cloud complex are also marked with open circles. The blue square outlines the region covered in the present work. The dashed line AB shows the position–velocity (PV) cut used to generate the $^{12}$CO PV diagram (Fig.~\ref{Fig10}). The scale bar corresponds to 50~pc, assuming a distance of 8.2 kpc to the GC.}
\label{Fig1}
\end{figure*}

Towards higher latitudes of this interaction zone, the Helix stream \citep{2024A&A...689A.121V} emerges as a striking molecular structure arching upward from the G1.3 cloud. This high-velocity stream exhibits a remarkable double-helix morphology extending over 200~pc, with gas velocities reaching up to V$_\mathrm{LSR} \sim$ +150--200\kms. Our previous work revealed extensive SiO emission along the Helix stream, indicative of widespread molecular shocks. Multi-wavelength analyses of \citet{2024A&A...689A.121V} further uncovered evidence of cloud-cloud collisions, localised star formation, and a feedback-driven expanding shell. Despite these insights, the region between the hypothesised dust lane impact point at G1.3 and the base of the Helix stream (Fig.~\ref{Fig1}) has remained largely unexplored. This intermediate zone likely marks the transition from inflow-dominated gas to stream material overshooting the Galactic plane, providing a promising site for tracing dynamic interactions such as shocks and cloud collisions.

Methanol (CH$_3$OH) is one of the most abundant complex organic molecules in the interstellar medium, exhibiting numerous transitions across the centimetre, millimetre, and sub-millimetre regimes, including prominent maser lines. Methanol masers are classified into two categories \citep{1991ASPC...16..119M}: Class~I, which are collisionally pumped and typically observed in shocked environments such as protostellar outflows and cloud–cloud collisions \citep[e.g.,][]{{1990ApJ...364..555P},{1992SvA....36..590S}}, and Class~II, which are radiatively pumped and closely associated with high-mass star forming regions \citep[e.g.,][]{{2001A&A...369..278M},{2010MNRAS.404.1029C},{2021A&A...651A..87O}}. Towards the GC region, \citet{2013ApJ...764L..19Y} reported 356 distinct 36.2 GHz Class~I maser spots within the inner $160\times43$~pc. The high methanol abundance in this region has been attributed to cosmic ray-induced photodesorption from methanol formed on ice mantles. A subsequent VLA survey covering $l\sim-0.6^\circ$ to $+0.4^\circ$ detected 2240 Class~I methanol masers with narrow line widths \citep[$\sim1$\kms;][]{2016ApJS..227...10C}. 

The first detection of Class~I masers outside the Milky Way was reported in NGC 253 \citep{2014ApJ...790L..28E}. Follow up studies \citep{{2017MNRAS.472..604E},{2022A&A...663A..33H}} found the maser emission in NGC 253 at the interface between the edge of the nuclear ring and both ends of the galactic bar, near the inner Lindblad resonance. The origin of masers was attributed to shocks driven by large-scale cloud-cloud collisions. Extragalactic detections of the 36.2~GHz class~I CH$_3$OH transition were also reported toward Maffei~2 and IC~342 \citep{2020A&A...633A.106H}. More recently, \citet{2026ApJ...999L..20C} reported the first 36.2~GHz class~I CH$_3$OH maser detection toward the barred spiral galaxy NGC~1365. The maser is localised to the southern bar-inflow lane, where no prominent star formation is observed. If bar-driven inflows interacting with the nuclear ring can trigger class~I methanol masers as observed in NGC 253 and NGC 1365, then the interaction between the G1.3 cloud and the base of the Helix stream provides a compelling Galactic analogue to test this scenario.

In this paper, we present the first observations of the 36.2~GHz $4_{-1}-3_0$~E and 44.1~GHz  $7_0-6_1$A$^+$ Class~I methanol maser transitions, along with the 48.4~GHz $1_0-0_0$A$^+$ thermal methanol line, towards the transitional region between the G1.3 cloud and the Helix stream, where bar-driven inflowing gas overshoots the CMZ and arches upward. We derive column densities and abundances of CH$_3$OH and SiO, a well-known shock tracer. By examining the spatial and kinematic relationship between maser emission and shock tracer, we investigate the origin of Class~I methanol masers in this region. The organisation of the paper is as follows: Details related to the data are given in Section 2. Section 3 describes the results of our multiwavelength analysis, whereas in Section 4 we discuss our findings. In Section 5 we present our conclusions.

\begin{figure*}[!htb]
\centering
\hspace*{-1.3cm}
\includegraphics[scale=0.45]{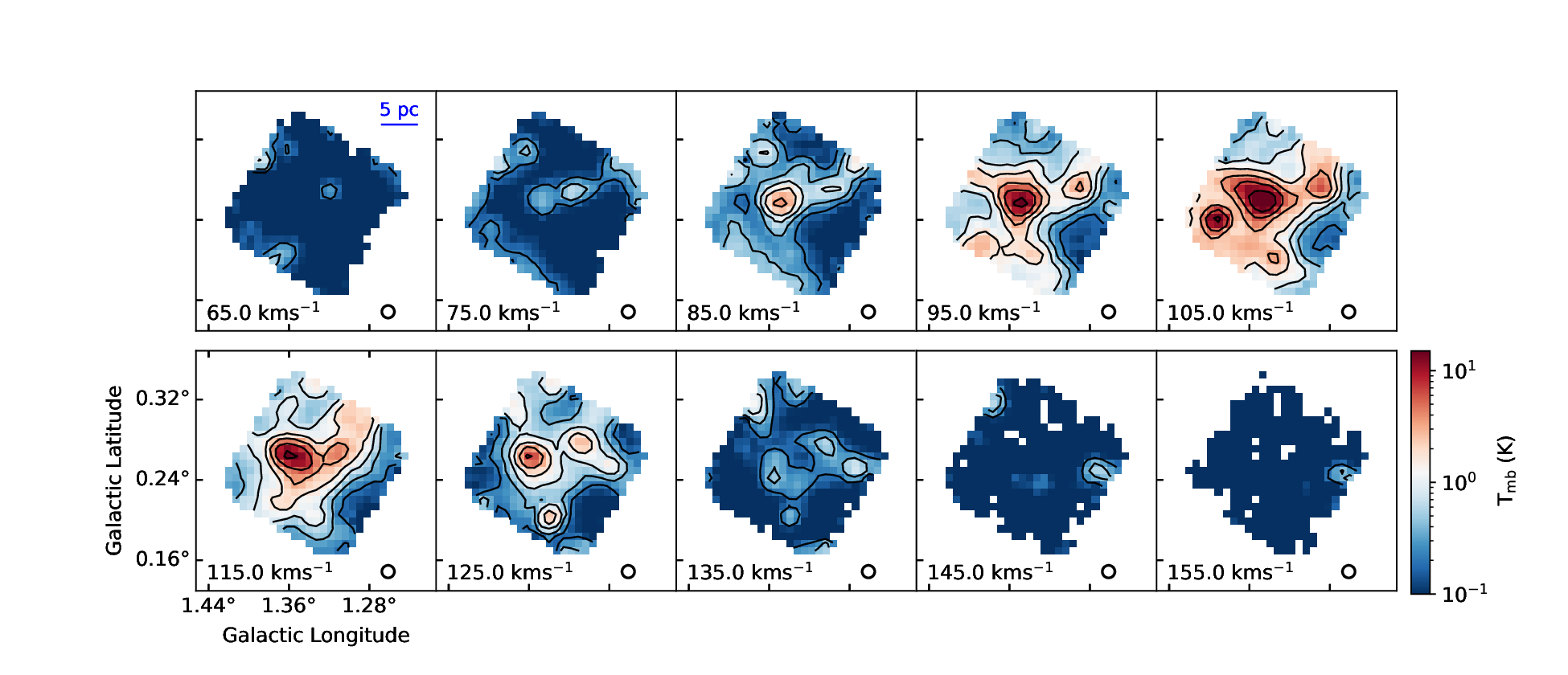}
\caption{Channel maps of CH$_3$OH emission at 36.2 GHz. Each panel shows the spatial distribution of emission at a velocity interval of 10 \kms, spanning 65 to 155 \kms. Contour levels are 0.2, 0.4, 0.8, 1.6, 3.2, 6.4, 12.8 and 25.6~K. The beam size is shown towards the bottom right of each panel.}
\label{Fig2}
\end{figure*}

\begin{figure*}[]
\centering
\hspace*{-1.3cm}
\includegraphics[scale=0.45]{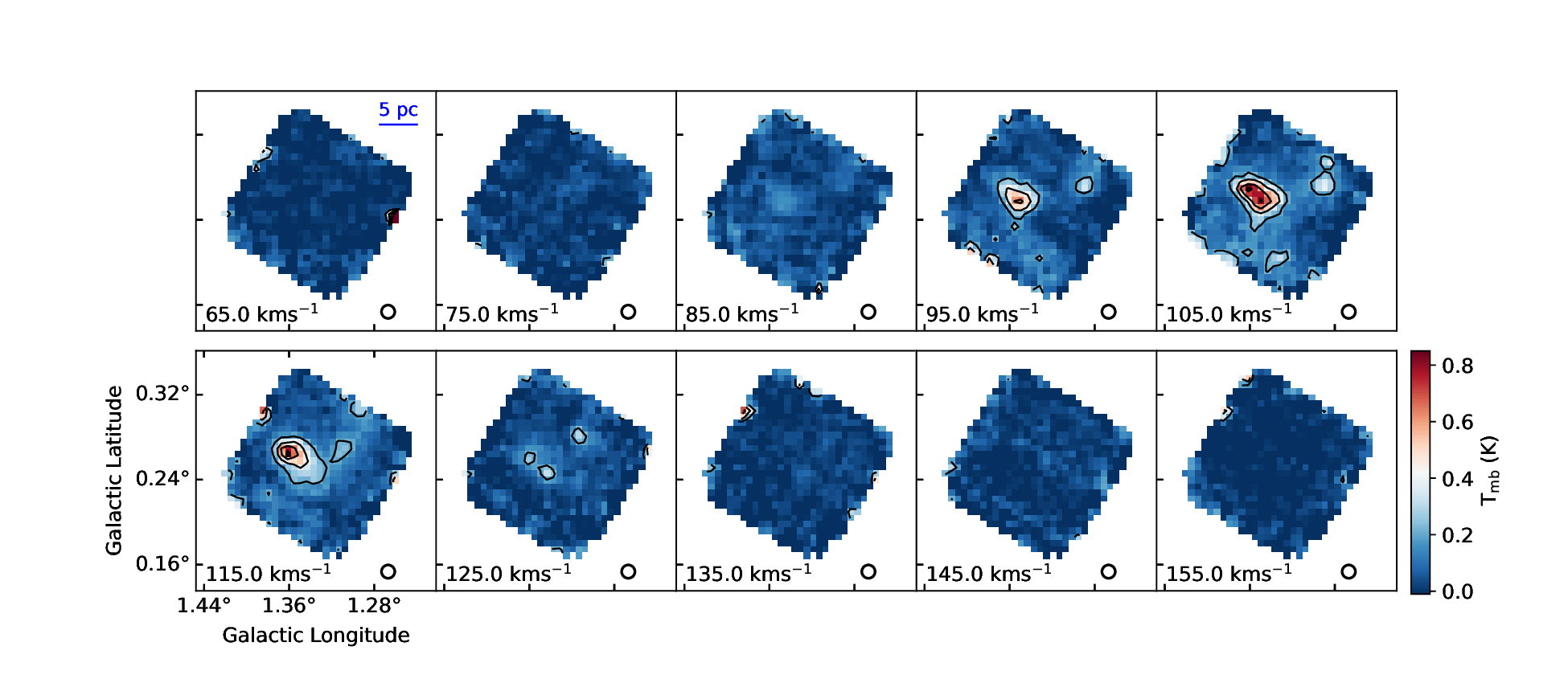}
\caption{Channel maps of CH$_3$OH emission at 44.1 GHz. Each panel shows the spatial distribution of emission at a velocity interval of 10 \kms, spanning 65 to 155 \kms. Contour levels are 0.2, 0.4, and 0.6~K. The beam size is shown towards the bottom right of each panel.}
\label{Fig3}
\end{figure*}

\begin{figure*}[]
\centering
\hspace*{-1.3cm}
\includegraphics[scale=0.45]{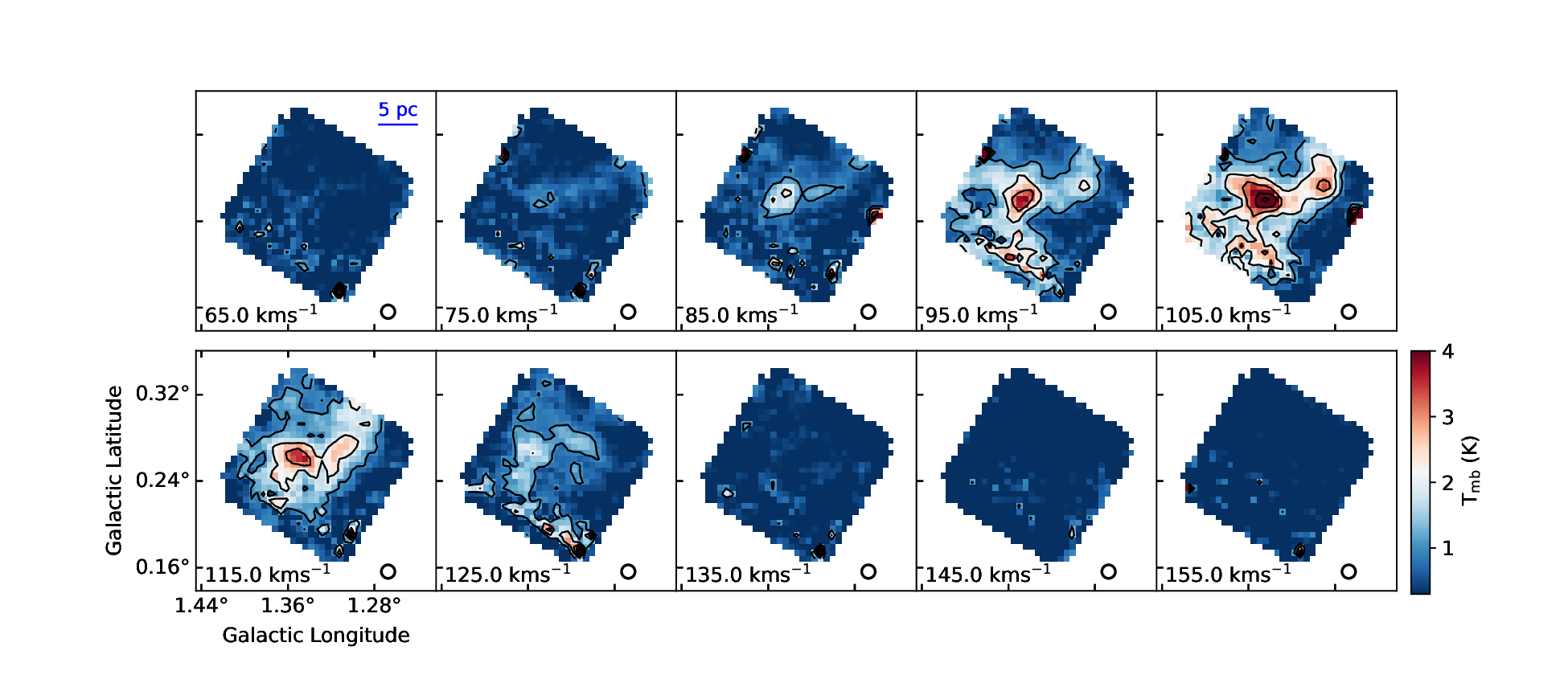}
\caption{Channel maps of CH$_3$OH emission at 48.4 GHz. Each panel shows the spatial distribution of emission at a velocity interval of 10 \kms, spanning 85 to 125 \kms.  Contour levels are from 1.0~K to 5.0~K, in steps of 1~K. The beam size is shown towards the bottom right of each panel.}
\label{Fig4}
\end{figure*}

\begin{table}[h]
\centering
\caption{Efficiencies, Jy/K conversion factors, and rms}
\setlength{\tabcolsep}{3pt}
\renewcommand{\arraystretch}{0.8}
\footnotesize
\begin{tabular}{c c c c c c}
\hline\\
Species & Frequency & Polarization& $\eta_{MB}^*$ & $Jy/K^*$ & rms \\
& (GHz) & & & & (K) \\
\hline\\
\multirow{2}{*}{CH$_3$OH $4_{-1}-3_0$~E} 
& \multirow{2}{*}{36.1692} & H& 0.61 & 3.95 & \multirow{2}{*}{0.07} \\
&  & V & 0.60 & 4.01 &  \\

\multirow{2}{*}{SiO ($J=1-0$)} 
& \multirow{2}{*}{43.4238} & H & 0.56 & 4.34 & \multirow{2}{*}{0.14} \\
&  & V & 0.55 & 4.38 &  \\

\multirow{2}{*}{CH$_3$OH $7_0-6_1$A$^+$} 
& \multirow{2}{*}{44.0694} & H & 0.56 & 4.34 & \multirow{2}{*}{0.20} \\
&  & V & 0.55 & 4.38 &  \\

\multirow{2}{*}{CH$_3$OH $1_0-0_0$A$^+$} 
& \multirow{2}{*}{48.3725} & H & 0.52 & 4.63 & \multirow{2}{*}{0.82} \\
&  & V & 0.51 & 4.73 &  \\
\hline
\end{tabular}
\label{Table1}
\vspace{1mm}
\begin{minipage}{\textwidth}
\footnotesize{$^*$Correspond to 2024 values (\url{https://rt40m.oan.es/rt40m_en.php})}
\end{minipage}
\end{table}

\section{Observations and archival data}
\subsection{CH$_3$OH and SiO observations}
To investigate the presence of potential Class~I methanol masers, we carried out 7 mm observations using the Yebes 40 m telescope between 11 November and 7 December 2024 (Project ID: 24B014). The observations utilized a HEMT receiver, which provides broad-band capabilities with an instantaneous bandwidth of 18 GHz (31.5--49.5~GHz) in two linear polarizations. Data were recorded with Fast Fourier Transform spectrometers (FFTs), each covering a 2.5 GHz bandwidth with a spectral resolution of 38 kHz \citep{2021A&A...645A..37T}. At these frequencies, the telescope’s half-power beam width (HPBW) ranges from 37$''$ to 59$''$. We performed On-The-Fly (OTF) mapping over a $9'\times9'$ region centred at $\alpha_{J2000}$: $17^\mathrm{h}47^\mathrm{m}46.9^\mathrm{s}$, $\delta_{J2000}$: $-27^\circ39'51.4''$ ($l=1.3341^\circ, b=0.2535^\circ$; centre of the blue square shown in Fig.~\ref{Fig1}). Data reduction was performed using the CLASS package of the GILDAS software \citep{pety2005_gildas}. From the wideband spectral data, we extracted $\pm$250\kms\, velocity windows centred around the 36.1692 GHz $4_{-1}-3_0$~E, 44.0694 GHz $7_0-6_1$A$^+$, and 48.3725~GHz $1_0-0_0$A$^+$  CH$_3$OH transitions. To investigate potential associations with shocks, we also extracted spectra around the SiO($J=1-0$) transition at 43.4238~GHz. The final data cubes were resampled to a spectral resolution of 0.8\kms\, for subsequent analysis. A first-order baseline subtraction was then applied to each spectrum to remove residual continuum or instrumental artifacts prior to line analysis. The spectra were initially measured in antenna temperature units and then converted to main beam brightness temperature ($T_\textrm{MB}$) using a forward efficiency of 0.97 and appropriate beam efficiency factors, while for the maser transitions we additionally express the intensities in Jy using the Jy/K conversion factors (Table~\ref{Table1}). Radial velocities are given on the Local Standard of Rest (LSR) scale. 

\subsection{Archival CO data from CHIMPS2 survey}
To trace the large-scale molecular gas kinematics of the region, we use $^{12}$CO($J=3-2$) data from the CO Heterodyne Inner Milky Way Plane Survey 2 \citep[CHIMPS2;][]{2020MNRAS.498.5936E}. CHIMPS2 is a large-scale CO survey of the Galaxy, covering the inner Galaxy, the CMZ, and part of the outer Galaxy, carried out with the 15~m James Clerk Maxwell Telescope (JCMT; angular resolution: 15$''$, velocity resolution: 1~\kms, rms noise: $\Delta T_{MB}$ = 0.81 K). The observational setup and data reduction are described in detail by \citet{2020MNRAS.498.5936E}. Reduced $^{12}$CO mosaics ($2^\circ\times1^\circ$) are publicly available in FITS format from the CANFAR archive \footnote{\url{https://www.canfar.net/citation/landing?doi=20.0004}}.

\begin{figure*}[!htb]
\centering
\includegraphics[scale=0.47]{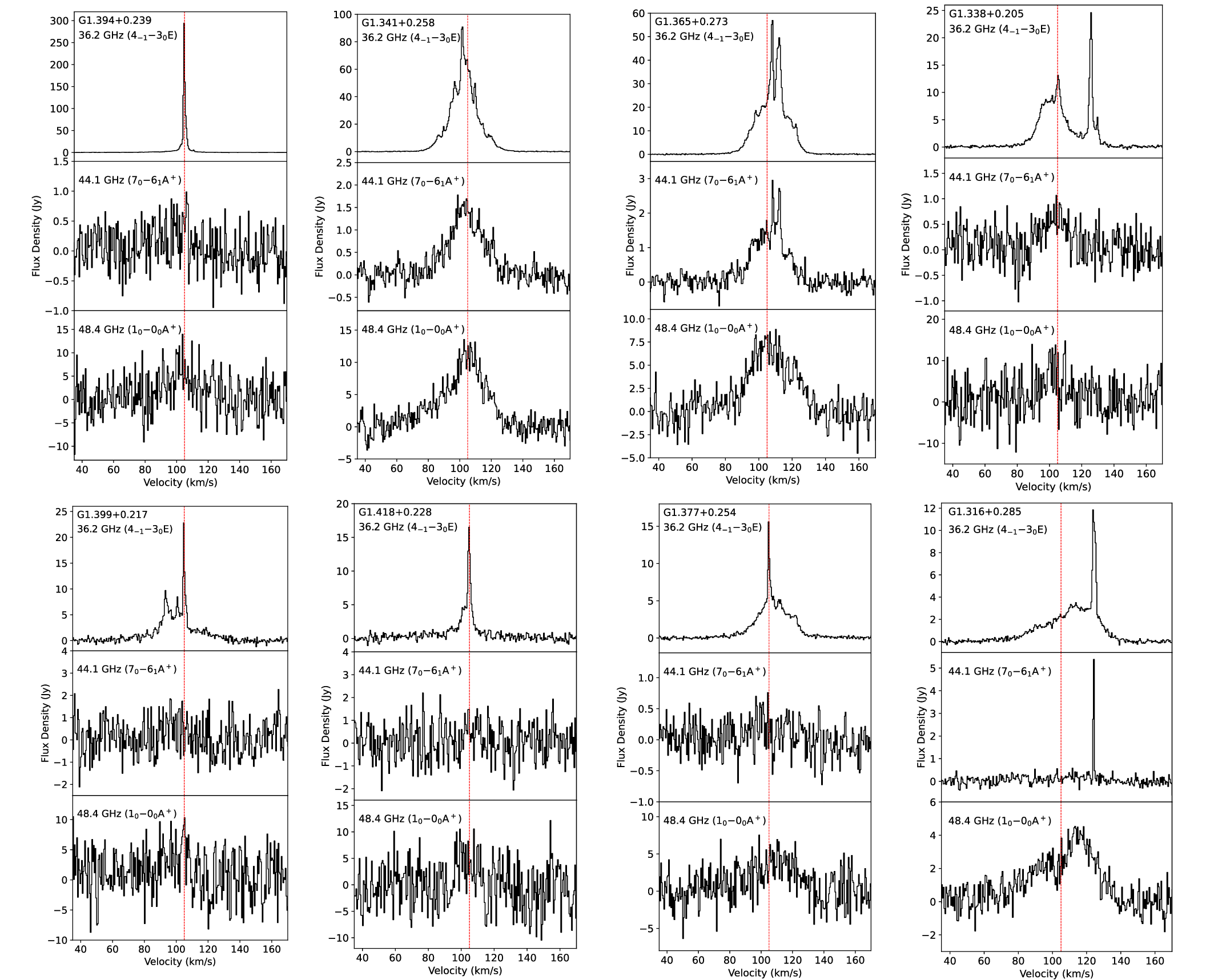}
\caption{Beam-averaged spectra of 36.2~GHz $4_{-1}-3_0$E (top panel), 44.1~GHz $7_0-6_1$A$^+$ (middle panel), and 48.4~GHz $1_0-0_0$A$^+$ (bottom panel) CH$_3$OH transitions toward positions that satisfy the maser candidate criteria (see Sect. 3.1 and Fig.~\ref{Fig7}). The vertical dotted line corresponds to V$_\textrm{LSR}$ = 105\kms.}
\label{Fig5}
\end{figure*}

\begin{figure*}[!htb]
\ContinuedFloat
\centering
\includegraphics[scale=0.47]{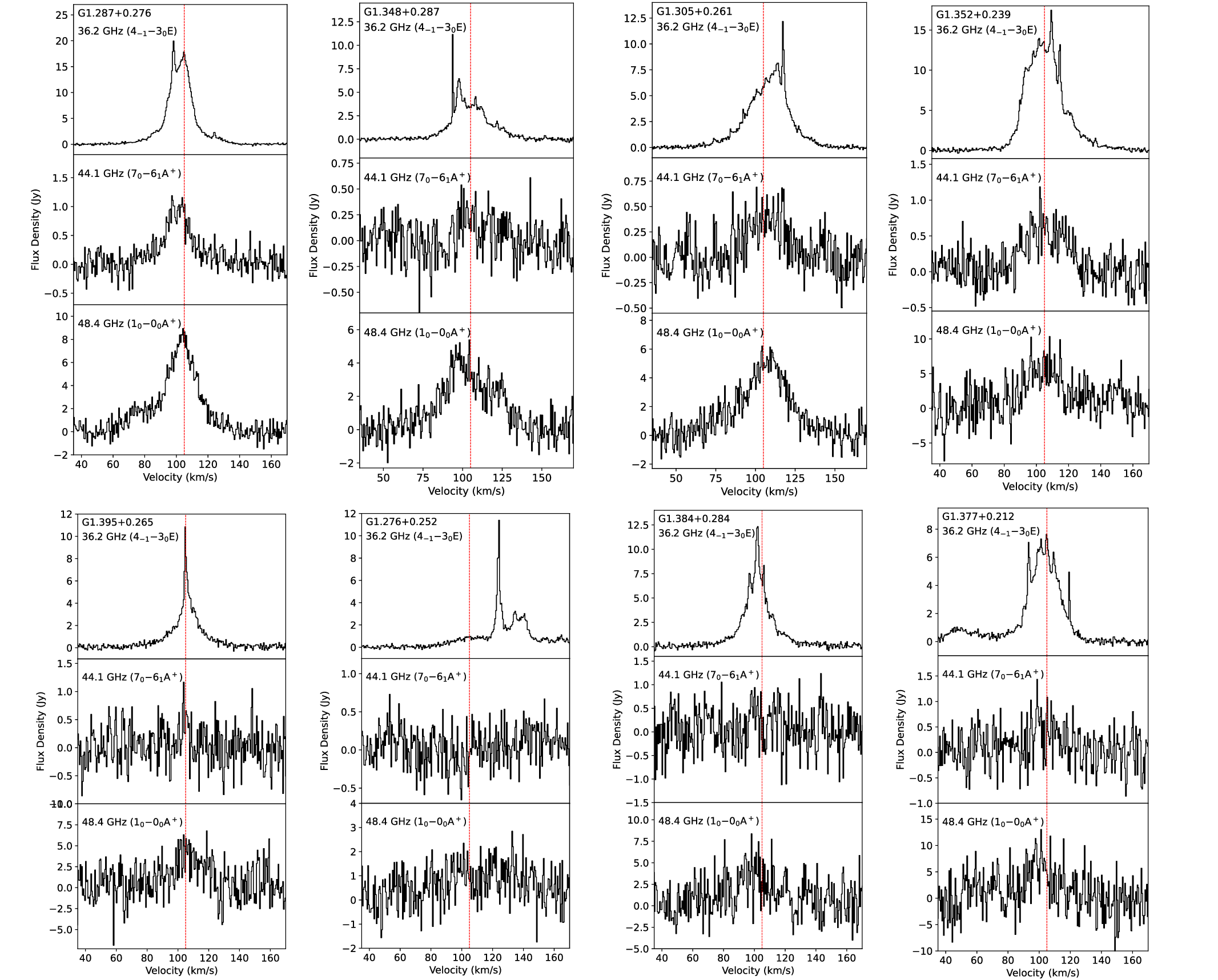}
\caption{(Continued.)}
\label{Fig5}
\end{figure*}

\section{Results}
The channel maps of the CH$_3$OH emission at 36.2~GHz, 44.1~GHz, and 48.4~GHz are presented in Figs.~\ref{Fig2}, \ref{Fig3}, and \ref{Fig4}. We detect widespread methanol emission in the region, extending over the entire mapped spatial scale of $\sim$24~pc. At 36.2~GHz, the emission spans a wide velocity range from 65 to 155\kms, and appears spatially extended, exhibiting multiple clumpy structures embedded within. This emission could be a combination of thermal and maser components. In contrast, the 44.1~GHz CH$_3$OH emission is detected over a narrower velocity range, from 95 to 125\kms, and is more compact in morphology. While some low-level extended emission is present, the dominant structures are compact peaks. This transition is known to exhibit maser as well as quasi-thermal emission \citep{1997ApJ...474..346M}. Notably, we observe a prominent emission peak at ($l,b$) = (1.3595$^\circ$, 0.2688$^\circ$), which spatially coincides with one of the brightest features in the 36.2~GHz channel maps, indicating a region of strong methanol emission at both frequencies. The morphology of 48.4~GHz thermal emission, corresponding to the ground state $1_0-0_0$A$^+$ transition, is broadly similar to the 36.2~GHz emission, although it misses some of the bright peaks seen in the 36.2~GHz maps (e.g., channel around V$_\mathrm{LSR}\sim$105\kms), which could likely arise due to maser emission.

\begin{table*}[!htbp]
\footnotesize
\caption{Spectrally decomposed CH$_3$OH maser line parameters at 36.2~GHz}
\setlength{\tabcolsep}{4pt}
\renewcommand{\arraystretch}{0.9}
\centering
\begin{tabular}{c c c c c c c c c c c c}
\hline\\
ID&Source&GLON$^1$&GLAT$^2$&$\textrm{S}_p$$^3$&$\textrm{V}_p$$^4$&$\Delta$V$^5$&$\int{S_{36.2}\,\textrm{dV}}^6$&L$_{36.2}^7$&$F_{36.2}^{\Delta v^8}$&$F_{44.1}^{\Delta v^9}$&$R^{10}_{36.2/44.1}$ \\
&&(Deg)&(Deg)&(Jy)&(\kms)&(\kms)&(Jy\kms)&L$_\odot$&(Jy\kms)&(Jy\kms)&  \\
\hline\\
1&G1.394+0.239&1.39401&0.23903&250.0&105.0&1.4&372.6&$0.9\times10^{-3}$&342.7&3.9&87.9\\
2&G1.341+0.258&1.34054&0.25755&40.8&101.7&1.6&69.4&$1.8\times10^{-4}$&967.2&40.0&24.2\\
&&&&22.7&109.9&2.1&50.7&$1.3\times10^{-4}$&&&\\
&&&&10.4&107.1&0.9&9.9&$2.5\times10^{-5}$&&&\\
3&G1.365+0.273&1.36536&0.27254&34.3&108.3&1.2&43.7&$1.1\times10^{-4}$&568.9&36.4&15.6\\
&&&&5.5&98.4&1.5&8.8&$2.2\times10^{-5}$&&&\\
&&&&4.3&122.6&1.4&6.4&$1.6\times10^{-5}$&&&\\
4&G1.338+0.205&1.33819&0.20533&21.8&125.9&1.9&44.1&$1.1\times10^{-4}$&200.6&12.3&16.3\\
&&&&5.2&105.7&2.1&11.6&$2.9\times10^{-5}$&&&\\

5&G1.399+0.217&1.39906&0.21672&18.6&105.1&1.0&19.8&$5.0\times10^{-5}$&127.6&2.8&45.6\\
6&G1.418+0.228&1.41764&0.22793&11.5&105.1&1.3&15.9&$4.0\times10^{-5}$&44.7&1.6&28.4\\
7&G1.377+0.254&1.37663&0.25389&9.8&105.1&1.2&12.5&$3.2\times10^{-5}$&141.7&7.7&18.4\\
8&G1.316+0.285&1.31568&0.28474&9.5&124.3&1.3&13.2&$3.3\times10^{-5}$&110.8&6.1&18.2\\
&&&&5.6&125.5&1.6&9.5&$2.4\times10^{-5}$&&&\\
9&G1.287+0.276&1.28711&0.27595&7.1&98.5&1.5&11.4&$2.9\times10^{-5}$&248.6&16.8&14.8\\
10&G1.348+0.287&1.34788&0.28734&6.5&94.2&0.9&6.2&$1.6\times10^{-5}$&115.0&6.7&17.2\\
11&G1.305+0.261&1.30449&0.26112&6.2&117.7&1.1&7.3&$1.8\times10^{-5}$&190.6&9.6&20.0\\
12&G1.352+0.239&1.35181&0.23889&6.2&114.9&1.7&11.2&$2.8\times10^{-5}$&320.7&20.3&15.8\\
&&&&4.3&109.6&1.3&5.9&$1.4\times10^{-5}$&&&\\
13&G1.395+0.265&1.39520&0.26511&5.9&105.1&1.2&7.5&$1.9\times10^{-5}$&82.9&3.4&24.4\\
14&G1.276+0.252&1.27601&0.25238&5.8&124.2&1.9&11.6&$2.9\times10^{-5}$&79.1&0.6&141.2\\
15&G1.384+0.284&1.38393&0.28377&5.2&102.1&1.8&9.9&$2.5\times10^{-5}$&145.3&7.4&19.7\\
16&G1.377+0.212&1.37678&0.21166&3.9&93.6&1.8&7.5&$1.9\times10^{-5}$&145.6&6.9&21.0\\
\hline\\
\end{tabular}
\vspace{1mm}
\begin{minipage}{\textwidth}
\footnotesize{$^1$Galactic longitude; $^2$Galactic latitude; $^3$Peak flux; $^4$Peak velocity; $^5$Line width; $^6$Integrated flux density; $^7$Maser luminosity; $^8$Velocity-integrated \\flux density at 36.2~GHz; $^9$Velocity-integrated flux density at 44.1~GHz; $^{10}$Ratio of the velocity integrated flux density ratios between 36.2 and 44.1 GHz } 
\end{minipage}
\label{Table2}
\end{table*}

\subsection{Maser identification using $\tt{GAUSSPY+}$}

The average spectra of CH$_3$OH emission within a circular aperture of radius 205$''$ at 36.2, 44.1, and 48.4~GHz are shown in Fig.~\ref{FigA1}. The 36.2~GHz ($4_{-1}-3_0$E) spectrum displays a broad, asymmetric emission profile with extended wings on both sides, suggestive of multiple blended components. The 44.1~GHz ($7_0-6_1$A$^+$) line, while weaker in intensity, also displays a broad profile. To identify potential Class~I maser emission, it is essential to decompose the spectra into individual components. We therefore performed a pixel-by-pixel spectral decomposition of the 36.2~GHz CH$_3$OH data using the $\tt{GAUSSPY+}$ package \citep{2015AJ....149..138L,2019A&A...628A..78R}. A similar decomposition was not performed for the 44.1~GHz data due to lower signal-to-noise ratios. $\tt{GAUSSPY+}$ is an extension of $\tt{GaussPy}$, a Python-based autonomous Gaussian decomposition algorithm designed to analyse complex interstellar medium spectra by fitting them with multiple Gaussian components. It employs derivative spectroscopy to automatically determine initial guesses for component parameters, and includes an automated spatial refitting stage that improves fits based on solutions from neighbouring pixels.

We adopted the default $\tt{GAUSSPY+}$ decomposition parameters by \citet{2019A&A...628A..78R}, except for the smoothing parameters $\alpha_1$ and $\alpha_2$. In particular, we used a minimum signal-to-noise threshold of S/N = 3 and a significance threshold of 5 for component identification. The default automated refitting options for blended components, broad features, negative residual peaks, and spatially inconsistent fits were retained. We found these parameter choices to be adequate for our data as the RMS noise properties across the mapped region were found to be relatively uniform, with increased noise primarily confined to the map edges. The smoothing parameters $\alpha_1$ and $\alpha_2$ were optimised using the training in $\tt{GAUSSPY+}$ on a set of 200 randomly selected spectra from the data cube, yielding $\alpha_1 = 1.33$ and $\alpha_2 = 6.63$. These smoothing parameters set the widths of the Gaussian kernels used to filter the spectra before the derivatives are calculated. This step suppresses noise peaks while preserving real emission features, ensuring stable identification of maxima and minima. $\alpha_1$ controls the detection of narrower components, while $\alpha_2$ introduces an additional smoothing scale needed when both narrow and broad line widths are present. A subset of 20 spectra was visually inspected to verify the quality of the resulting decomposition. The inspection focused on whether the fitted Gaussian components adequately reproduced the observed line profiles, whether significant residuals remained after subtraction of the fitted model, and whether any clearly spurious narrow components were introduced. We examined the residual spectra returned by the decomposition, together with the measured noise level and the reduced chi-square values. For most spectra, the residuals were generally consistent with the noise, with no systematic excess features remaining after the fits. The resulting decompositions reproduced both broad blended emission and narrow spectral features sufficiently well for the purpose of identifying candidate narrow components. Following the initial decomposition, the subsequent automated quality-control and refitting steps implemented within $\tt{GAUSSPY+}$ were applied to account for  blended components, broad profiles, non-Gaussian residuals, components inconsistent with neighbouring fits, and discontinuities across adjacent spectra. In the resulting decomposition, we found up to 16 components toward the pixel with the maximum integrated CH$_3$OH intensity, with an average of 7 components per pixel across the region. In total, this yielded 1822 Gaussian components across the mapped area.

The multi-component Gaussian fit reproduces the observed line profiles well, with minimal residuals. The line profiles range from broad, blended features to narrow, isolated peaks. A subset of pixels display strong, narrow features indicative of potential maser emission. The broadest line has a full width at half maximum (FWHM) $\Delta$V = 36.9\kms, whereas the median FWHM is 3.0\kms. To isolate potential maser candidates from these, we applied two selection criteria: (1) a line width FWHM ($\Delta\mathrm V$) less than 2.35\kms, and (2) a conservative signal-to-noise ratio (S/N) greater than 30. This resulted in 192 components that exhibit characteristics consistent with maser emission. To avoid duplication due to spatial overlap within the telescope beam, we identified unique maser positions by retaining only the brightest (peak) pixel within a 75$''$ radius ($\sim1.4\times\theta_{36.2}$). Applying this spatial filtering, we arrive at a final set of 16 spatially distinct maser candidates. Several of these exhibit multiple velocity components, resulting in a total of 23 components across all positions. To test the sensitivity of our choice of S/N threshold, we repeated the selection using a lower threshold of S/N $>$ 20 while keeping the line width criterion unchanged. This increased the number of selected narrow Gaussian components from 192 to 251. However, after applying the same 75$''$ spatial filtering, the number of spatially distinct maser candidates changed only from 16 to 17.  The additional candidate lies near the edge of the mapped emission and does not affect the main conclusions. Our final catalogue is therefore not strongly sensitive to the adopted S/N threshold. To ensure that our results are robust to spectral resolution, we also tested the decomposition using a finer velocity resolution of 0.5\kms. As expected, this resulted in a greater number of Gaussian components. However, after applying the same spatial filtering procedure and grouping multiple velocity components associated with the same maser position, the number of unique maser candidates remained effectively unchanged. This confirms that the 0.8\kms\ resolution preserves all relevant physical structure while avoiding overfitting and ensuring more stable component fitting. 

The peak spectra of these maser candidates, along with their corresponding $\tt{GAUSSPY+}$ decomposed Gaussian components and residuals are presented in Fig.~\ref{FigA2}. While the multi-component Gaussian fits broadly reproduce the observed line profiles, a number of maser candidates exhibit noticeable residuals, likely due to the narrow, high-intensity, and potentially non-Gaussian nature of maser emission. These residuals highlight the limitations of purely Gaussian decomposition of such features, though the overall decomposition remains informative in complex spectral environments. 

\subsection{Properties of maser candidates}

The Gaussian decomposition performed using $\tt{GAUSSPY+}$ indicates that the 36.2~GHz $4_{-1}-3_0$E CH$_3$OH spectra toward 16 independent positions satisfy the maser candidate criteria. These spectra typically exhibit a combination of broad, low-amplitude components consistent with thermal/quasi-thermal emission and narrow, high-intensity features characteristic of maser amplification. Fig.~\ref{Fig5} presents the beam-averaged spectra extracted from the 36.2, 44.1, and 48.4~GHz data cubes at the candidate positions. Overall, the 44.1~GHz emission is 1--2 orders of magnitude weaker than the 36.2~GHz emission. While faint 44.1~GHz emission is present toward most positions, it is generally broad and of low intensity compared to the 36.2~GHz line. Narrow, maser-like 44.1~GHz features are identified only toward G1.365+0.273 and G1.316+0.285; similarly narrow features in other spectra occur at amplitudes comparable to the rms noise and are therefore not considered significant. We compute the flux ratio, $R_{36.2/44.1}$, by integrating the 36.2 and 44.1 GHz flux densities over the 36.2 GHz beam and across the same velocity range. The resulting ratios in our region lie in the range 14.8--141.2 (Table~\ref{Table2}). This differs from the trend reported by \citet{2014MNRAS.439.2584V} in Galactic star-forming regions, where the 44.1~GHz line is generally brighter than the 36.2~GHz line, with only $\sim14\%$ of maser groups exhibiting stronger 36.2~GHz emission. They also note that due to spectral complexity, this percentage likely represents an upper limit. 

The prevalence of maser and strong thermal features in the 36.2~GHz line in our region of interest, in contrast to the lack of or comparatively faint emission in the 44.1~GHz line under identical observing conditions, likely reflects differences in excitation requirements. Such conditions have been linked to environments with little or no active star formation \citep{2008AJ....135.1718P}, or to shock-driven regions such as supernova remnants where 36.2~GHz masers have been observed in the absence of 44.1~GHz emission \citep{2014AJ....147...73P}. We will discuss this in detail in Section 4. In addition to the methanol maser transitions, we extracted the 48.4~GHz CH$_3$OH line ($1_0$–$0_0$ A$^+$), which is expected to trace thermal emission, toward each candidate position. As shown in the bottom panels of Fig.~\ref{Fig5}, this transition generally appears as a broad, low-intensity feature, consistent with thermal or quasi-thermal excitation.

\begin{figure}[!thbp]
\centering
\includegraphics[scale=0.43]{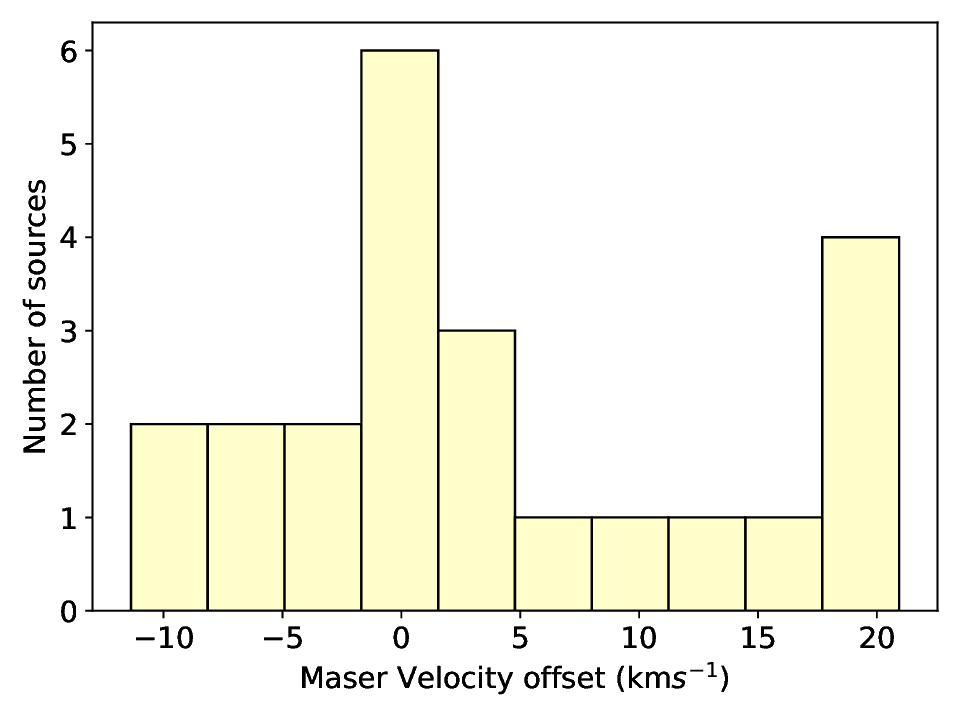}
\caption{Histogram of velocity offset distribution with respect to the LSR velocity of 105\kms\ for 23 maser components.}
\label{Fig6}
\end{figure}

The line parameters of identified maser positions are presented in Table~\ref{Table2}. Since the masers are point-like at our resolution, we extract the line parameters from the peak-pixel spectrum. All maser fluxes and integrated intensities in Table 2 are reported in Jy and Jy\kms. Maser velocities range from 93.6\kms~to 125.9\kms and the line widths range between 0.9 and 2.1\kms. Integrated intensities span nearly two orders of magnitude, from 5.9 Jy\kms~to 372.6 Jy\kms. Of the 16 positions, 4 exhibit multiple velocity components. Six of the 23 maser components are found clustered around the LSR velocity of 105\kms (zero velocity offset), evident from the histogram of maser velocities presented in Fig.~\ref{Fig6}. There is a second clustering around velocity offset of 20~\kms (4 components). Among the 23 maser lines, 15 (65$\%$) are within $\pm10\kms$ of the LSR velocity of 105\kms. Two particularly interesting maser sites in our sample, G1.365+0.273 and G1.338+0.205, exhibit multiple maser components with large velocity separations. G1.365+0.273 hosts three distinct maser components at 98.4, 108.3, and 122.6\kms, spanning over 24\kms. Similarly, G1.3382+0.2053 shows two components at 105.7 and 125.9\kms, separated by more than 20\kms. The isotropic luminosity of the 36.2~GHz masers can be calculated using the equation

\begin{equation}
\mathrm{L}_{36.2}=3.77\times10^{-8}\,\left(\frac{d}{1\,\mathrm{kpc}}\right)^2\,\left(\frac{\int S_{36.2}\mathrm{dV}}{1\,\mathrm{Jy}\,\kms}\right)   \mathrm{L}_\odot
\end{equation}

\begin{figure*}[!thbp]
\centering
\includegraphics[scale=0.28]{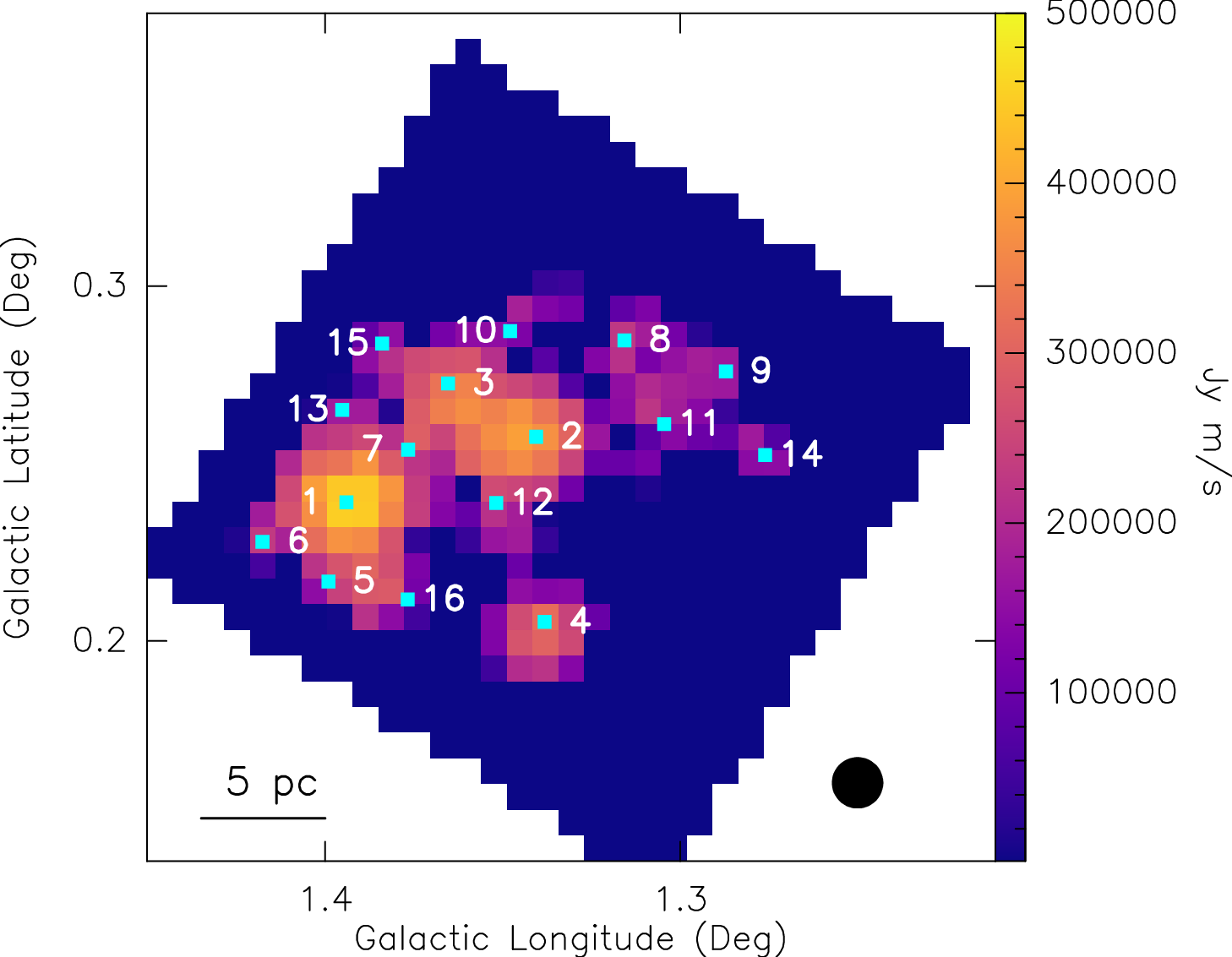}\quad \includegraphics[scale=0.28]{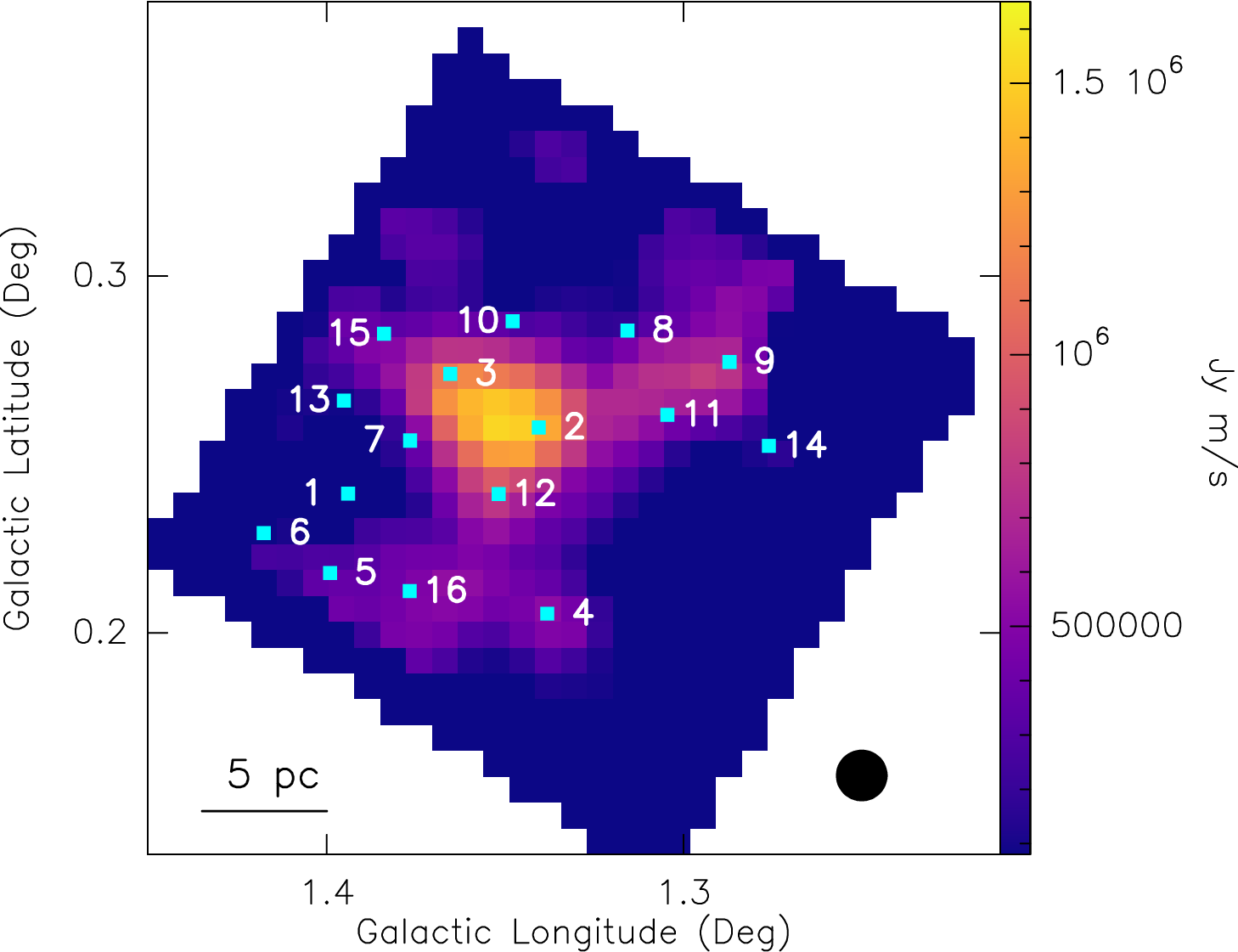} 
\caption{(Left) Integrated intensity emission at 36.2~GHz from all pixels satisfying the maser criteria. (Right) Integrated intensity map of the 36.2~GHz thermal emission reconstructed from the Gaussian components identified as thermal in the decomposition. The positions of the 16 detected masers are marked and labelled with numbers corresponding to their IDs in Table~\ref{Table2}.}
\label{Fig7}
\end{figure*}

\noindent where d is the distance to the source \citep[assumed to be 8.2~kpc;][]{2019A&A...625L..10G}, and $\int S_{36.2}\textrm{dV}$ is the integrated flux density in units of Jy\,\kms. The numerical coefficient $3.77\times10^{-8}$ incorporates the conversion from Jy\kms\, to luminosity units using the rest frequency of the 36.2 GHz and expresses the result in units of $\mathrm{L}_\odot$. The estimated maser luminosities are listed in Table~\ref{Table2} and ranges from $1.4\times10^{-5}$~L$_\odot$ to $0.9\times10^{-3}$~L$_\odot$ . Comparison with Class~I masers associated with star forming regions and the GC region \citep[e.g.,][]{{2011ApJS..196...21B},{2019ApJS..244....2K},{2022ApJ...936..186B},{2023A&A...675A.112Y}} shows that the brightest 36.2~GHz maser in our sample, G1.394+0.239, has an estimated isotropic luminosity of $0.9\times10^{-3}$ L$_\odot$, placing it among the most luminous Class~I CH$_3$OH masers known in the Galaxy. 
  
\subsection{Extended maser and shocked SiO emission}

To further investigate the spatial distribution and physical origin of the 36.2~GHz CH$_3$OH emission, we performed a component-wise separation of maser and thermal emission using the fitted Gaussian parameters from the $\tt{GAUSSPY+}$ decomposition. Each pixel in the data cube was analysed individually, and spectral components were classified as maser-like or thermally-dominated, based on criteria specified in Section 3.1. With this pixel-by-pixel analysis, we constructed two synthetic cubes: one representing the maser-dominated emission, and the other the thermally-dominated emission. Integrated intensity maps of the resulting 36.2~GHz maser and thermal data cubes are presented in Fig.~\ref{Fig7}. As evident from Fig.~\ref{Fig7}(Left), we find extended maser emission across the region, with the brightest maser G1.394+0.239 located toward the south-east ($\Delta l=3.6'$,  $\Delta b=-0.9'$ relative to the map centre). Although 16 distinct maser spots were identified earlier, the morphology suggests that several of them are part of larger maser complexes extending over $\sim$5~pc. The thermally-dominated emission, shown in panel (b), is also spatially extended, but its peak is offset from the maser peak. The second brightest maser in the region, G1.341+0.258, is found adjacent to the thermal peak. Additionally, the thermal emission appears elongated along the northwest–southeast direction. To quantify the spatial extent of the brightest Class~I maser G1.394+0.239, we fitted a two-dimensional Gaussian to a $10\times10$ pixel cutout of the integrated intensity map. In Fig.~\ref{FigA3} we present the integrated intensity cutout (left), the best fit Gaussian model (middle), and the residual map (right). The best-fit convolved source size is $65.4''\times54.1''$ and after deconvolving the $52''$ beam, the intrinsic source size is $39.7''\times14.9''$ ($1.6\times0.6$~pc$^2$ at the GC distance). Although the 2D Gaussian serves as a useful first-order approximation, the residuals exhibit coherent structure rather than noise-like scatter, suggesting that the maser emission includes substructure not accounted for by a simple Gaussian model.

Since Class~I methanol masers are collisionally pumped and often correlate with molecular shock tracers \citep[e.g.,][]{2004ApJS..155..149K}, we used the SiO (1--0) line at 43.4~GHz, to investigate their association with shocks. Interstellar SiO, a robust tracer of shocked gas, is produced through the sputtering of Si-bearing material from dust grains \citep{1997A&A...321..293S}. The channel maps of SiO (1--0) emission are presented in the appendix, in Fig.~\ref{FigA4}. We detect large-scale SiO emission across the entire mapped region, spanning $\sim24$~pc, similar to 36.2 and 48.4~GHz CH$_3$OH emission and in the velocity range $\sim$75--145\kms. In order to compare the SiO and 36.2~GHz CH$_3$OH emission, we present a colour composite image in Fig.\ref{Fig8}, where maser emission is shown in red, thermally dominated CH$_3$OH emission in blue, and SiO emission as contours. The peak of the SiO emission coincides with the thermally dominated emission peak, with a morphology closely tracing that of the thermal CH$_3$OH emission. By contrast, the brightest maser-dominated feature toward the south-east peaks in a region of comparatively lower, but still detected, SiO emission. This is not unexpected, since SiO traces regions where shocks are sufficiently energetic to release silicon-bearing material from grains, whereas Class~I CH$_3$OH masers require favourable post-shock density, temperature, methanol abundance, and velocity-coherence conditions for collisional pumping. Thus, the lower SiO intensity at the maser peak does not rule out a shock origin, but may indicate that the maser-dominated and SiO-bright gas trace different parts (e.g. distinct shocks and/or post-shock conditions) of the same large-scale, turbulent interaction. At the current spatial resolution ($\sim1.7$~pc), the SiO spectra exhibit broad line profiles ($\Delta V \sim 25$–$35$~\kms), although this width is degenerate between a single high-dispersion component and a blend of unresolved narrow components.

\subsection{Column densities and abundances}

We estimate CH$_3$OH and SiO column densities under the assumptions of optically thin emission and local thermodynamic equilibrium (LTE). For the CH$_3$OH column density estimation, we use the 48.4~GHz 
ground-state transition, rather than the 36.2~GHz transition, in order 
to avoid contamination from maser and quasi-thermal emission. As an empirical check, we inspected the beam-averaged CH$_3$OH 48.4 GHz and SiO(1--0) spectra toward the 36.2~GHz thermal CH$_3$OH peak (Fig.~\ref{FigA5}). Both profiles are single-peaked and do not show obvious signatures of strong optical depth, such as pronounced self-absorption or flat-topped line profiles. Column densities are calculated using the following equation \citep{2023A&A...679A.123K}

\begin{equation}
\begin{split}
N_{tot} =\left( \frac{8\pi\nu^3}{c^3A_{ul}}\right) \left(\frac{Q({T})}{g_u}\right) \frac{\exp\left(\frac{E_u}{k_BT\mathrm{_{ex}}}\right)}{\exp\left(\frac{h\nu}{k_BT\mathrm{_{ex}}}\right)-1}\frac{1}{[J_\nu(T\mathrm{_{ex}})-J_\nu(T\mathrm{_{bg}})]}\\ \times \int\frac{T\mathrm{_{MB}}}{f}\,dV
\end{split}
\end{equation}

\noindent where $\nu$ is the frequency, $A_{ul}$ is the Einstein coefficient for spontaneous emission, $Q(\mathrm{T})$ is the partition function, $g_u$ is the statistical weight of the upper level, and $E_u$ is the energy of the upper level (see Table~\ref{Table3}). The molecular parameters for the CH$_3$OH $1_0-0_0$ A$^+$ and the SiO (1--0) lines are taken from the Cologne Database for Molecular Spectroscopy \citep[CDMS;][]{2005JMoSt.742..215M}. The excitation temperature $T_\textrm{ex}$ is assumed to be 20~K, based on the Hi-GAL dust temperature map derived using the point process mapping (PPMAP) tool \citep{2017MNRAS.471.2730M} available through the ViaLactea database\footnote {\url{http://www.astro.cardiff.ac.uk/research/ViaLactea/}}. The partition function at 20~K and 50~K are interpolated from CDMS tabulated values. The term $\int T\mathrm{_{MB}}\,dV$ represents the velocity-integrated main beam brightness temperature, and $f$ is the beam filling factor. Since the source is extended, we assume $f = 1$ and calculate beam-averaged column densities. The radiation temperature is defined as $J_\nu({T})\equiv\frac{h\nu/k_B}{\exp(h\nu/k_BT)-1}$, and $T_{bg}$ is the cosmic microwave background temperature, taken as 2.7~K.

\begin{table}[h!]
\centering
\footnotesize
\caption{Spectroscopic parameters used for the column density calculations.}
\begin{tabular}{lcc}
\hline
Parameter & CH$_3$OH (48.4 GHz) & SiO (43.4 GHz) \\
\hline
$A_{ul}$ (s$^{-1}$) &$3.5\times10^{-7}$  &$3.1\times10^{-6}$  \\
$g_u$              &3  & 3 \\
$E_u$ (K)          &2.3  &2.08  \\
$Q(T=20~\mathrm K)$             &85.75  & 19.53 \\
$Q(T=50~\mathrm K)$             &401.6  & 48.3 \\
\hline
\end{tabular}
\label{Table3}
\end{table}

\begin{figure}[!htbp]
\centering
\hspace*{-0.3cm}
\includegraphics[scale=0.28]{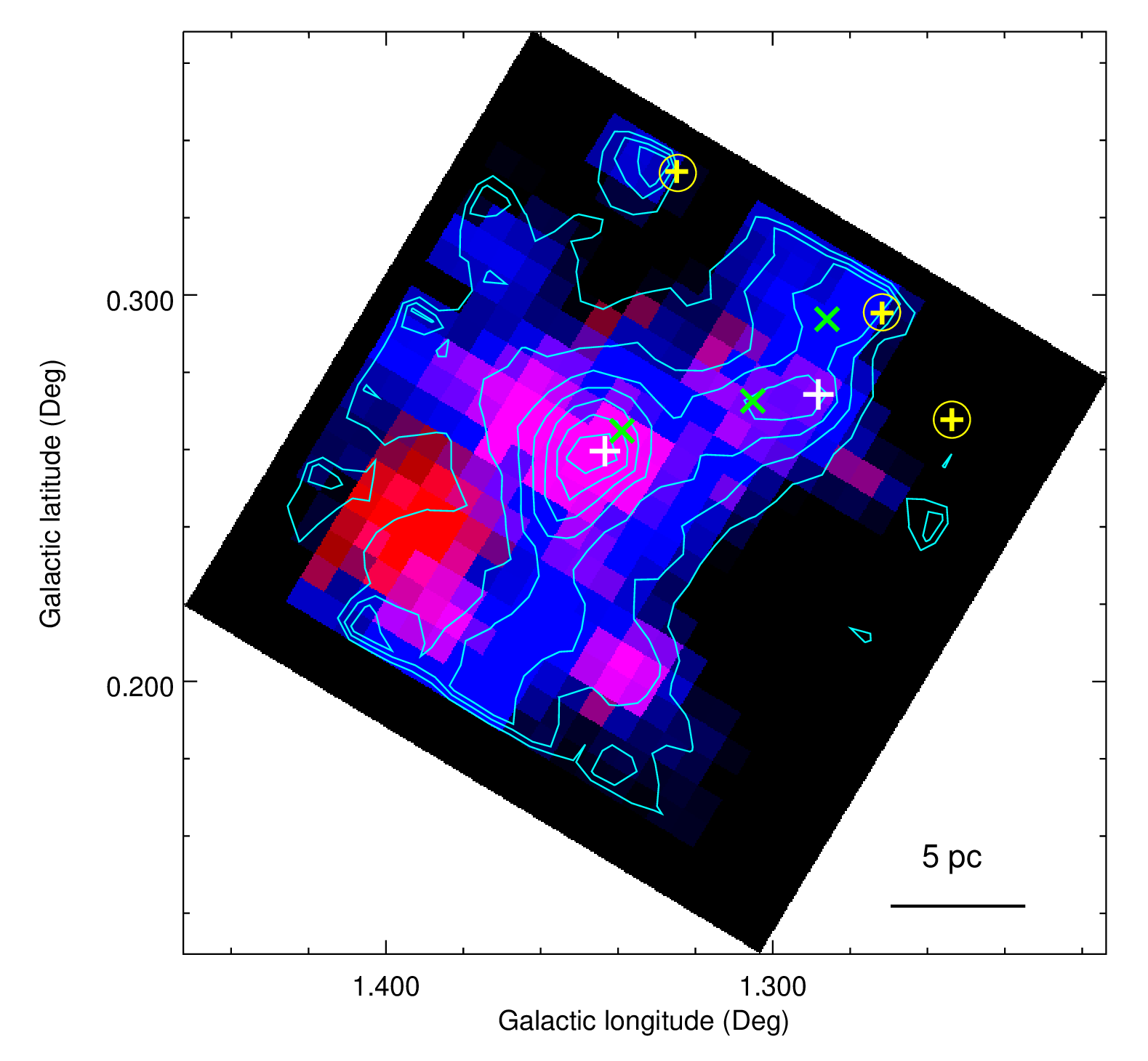}
\caption{Colour composite showing 36.2~GHz maser-dominated emission in red and 36.2~GHz thermally-dominated emission in blue. SiO integrated intensity emission in the velocity range 20\kms\, to 200\kms\, is overlaid as contours with levels from 23~K\kms\, to 73~K\kms\, in steps of 10~K\kms. White crosses indicate Hi-GAL prestellar clumps, green $\times$ mark protostellar clumps, and yellow encircled crosses denote prestellar clumps without reliable velocity and distance estimates.}
\label{Fig8}
\end{figure}

Using the CH$_3$OH and SiO(1--0) integrated intensity maps, we generated $N$(CH$_3$OH) and $N$(SiO) maps, masking emission below the $3\sigma$ level. For a direct comparison, the CH$_3$OH 48.4~GHz map was convolved and regridded to match the beam size and pixel scale of the SiO(1--0) map. The peak CH$_3$OH column density is $6.3\times10^{16}$~cm$^{-2}$ and the mean is $2.3\times10^{16}$~cm$^{-2}$, whereas for SiO the corresponding values are $8.9\times10^{14}$~cm$^{-2}$ and $3.1\times10^{14}$~cm$^{-2}$, respectively. We note, however, that this assumption does not necessarily imply $T_\mathrm{gas}=T_\mathrm{dust}$. In the Galactic centre, gas kinetic temperatures are often substantially higher than the dust temperature \citep[$T_\mathrm{gas}\sim$60~K to $>$100~K;][]{2016A&A...586A..50G}, and the observed emission may arise from shocked or post-shock gas in which grain processing has enhanced the gas-phase abundances of these species, while the gas has already cooled relative to the immediate shock front. Adopting $T_\mathrm{ex}=50$~K increases the mean column densities to $9.4\times10^{16}$~cm$^{-2}$ for CH$_3$OH and $6.7\times10^{14}$~cm$^{-2}$ for SiO. This demonstrates that single-line LTE estimates, especially for CH$_3$OH, may substantially underestimate the true column densities if the excitation temperature is higher than assumed. A more reliable determination would require a rotation-diagram or multi-line excitation analysis, which is not possible here because only ground-state transitions are available. We also note that there is 48.3769~GHz CH$_3$OH $1_0-0_0$ E line blue-shifted 
$\sim$27~\kms\ from the 48.3725~GHz $1_0-0_0$ A$^+$ line. Given the broad line widths in the region, the two transitions may be blended in some 
positions. Therefore, the integrated intensity of the 48.4~GHz feature 
may include a contribution from the E component, particularly on the 
blue-shifted side of the A$^+$ profile. Since the spectra do not show 
a clearly separated second peak over most of the mapped region, we do 
not attempt to decompose the A$^+$ and E contributions.

\begin{figure}[]
\centering
\hspace*{-0.3cm}
\includegraphics[scale=0.5]{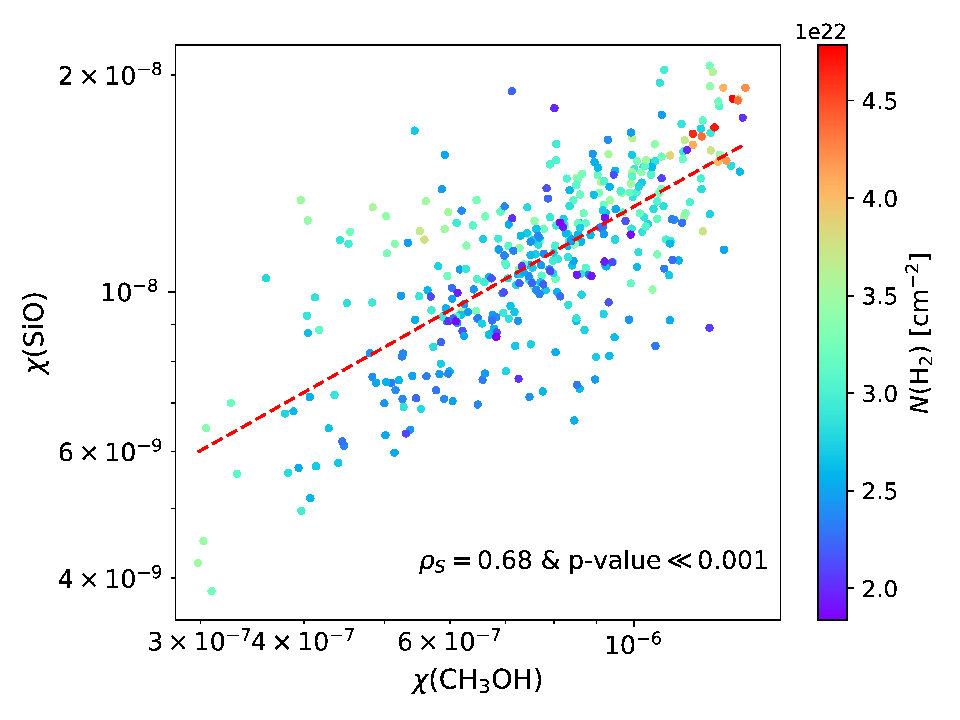}
\caption{Pixel-by-pixel comparison between CH$_3$OH and SiO fractional abundances (for $T_\mathrm{ex}=20$~K), both plotted in log–log space. The red dashed line indicates the best-fit power-law relation. A positive correlation is present (Spearman $\rho_S = 0.68$, $p \ll 0.001$), suggesting that the relationship between the two species is not solely driven by variations in total H$_2$ column density. Points are color-coded by $N$(H$_2$) to illustrate any dependence on gas column density.}
\label{Fig9}
\end{figure}

We define the fractional abundance of a species as $\chi$(X) = $N$(X)/$N$(H$_2$), where $N$(X) is the column density of the species and $N$(H$_2$) is the molecular hydrogen column density. For the abundance calculations, we used the molecular hydrogen column density map derived from the Hi-GAL dust continuum emission using the PPMAP tool. The PPMAP-based $N$(H$_2$) map was convolved and regridded to match the resolution and sampling of the SiO(1--0) map, enabling a consistent pixel-by-pixel comparison for the abundance calculations. Under the $T_\mathrm{ex}=20$~K assumption, the peak fractional abundance of CH$_3$OH is $1.4\times10^{-6}$ with a mean value of $8.0\times10^{-7}$, while for SiO the corresponding values are $2.1\times10^{-8}$ and $1.1\times10^{-8}$, respectively. For a higher excitation temperature of 50~K, the mean abundances increase to $\sim3.3\times10^{-6}$ for CH$_3$OH and $\sim2.4\times10^{-8}$ for SiO. The pixel-by-pixel comparison between $\chi$(CH$_3$OH) and $\chi$(SiO) (Fig.~\ref{Fig9}) reveals a positive monotonic correlation (Spearman $r_s=0.68$, $p\ll0.001$), indicating that the relationship between the two species is not solely an artifact of variations in total gas column density. Although the RGB composite in Fig.~\ref{Fig8} suggests that the maser-dominated emission (red) at G1.394+0.239 is associated with relatively weak SiO emission (contours), the fractional abundance ratio $\chi$(CH$_3$OH)/$\chi$(SiO) is $\sim$94, compared to $\sim$70.8 at the thermal CH$_3$OH peak. Thus, SiO is still present but at roughly half the abundance measured at the thermal peak.

We note, however, that both the column densities and fractional abundances are subject to systematic uncertainties arising from assumptions on excitation temperature, dust properties, and source filling factors. In particular, the $N$(H$_2$) values from PPMAP, while benefiting from improved spatial resolution and multi-temperature decomposition, carry uncertainties related to dust opacity, line-of-sight temperature structure, and background subtraction. Enhanced methanol abundances of $10^{-8}$–$10^{-6}$ are observed toward the G1.6–0.025 cloud, where \citet{2009ApJ...692...47M} identified signatures of a cloud–cloud collision. In that region, the CH$_3$OH emission is several times stronger than SiO, similar to what we observe in our target region. The enhanced abundances of both SiO and CH$_3$OH are interpreted as the result of shock chemistry. This case is particularly relevant to our study, as G1.6 together with G1.3 and our target field belong to the larger region at the edge of the CMZ where the inflowing bar-driven gas overshoots the CMZ. Under such conditions, Si locked in grain cores can be released into the gas phase by sputtering in magnetohydrodynamic C-shocks, followed by rapid oxidation to SiO \citep[e.g.][]{1997A&A...321..293S}. 

\begin{figure*}[!htbp]
\centering
\includegraphics[scale=0.45]{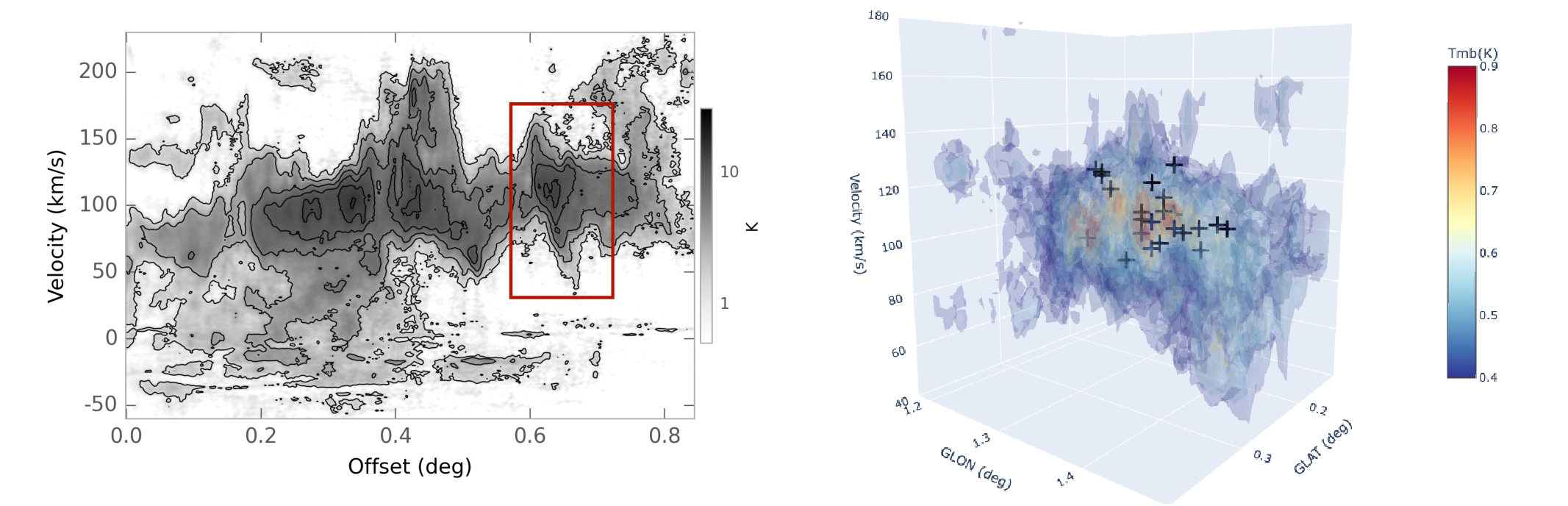}
\caption{(Left) $^{12}$CO(3--2) PV diagram from the CHIMPS2 survey. The rectangle marks the region investigated in the present study. (Right) PPV visualisation of the rectangular region highlighted in the left panel, with the positions of Class~I methanol masers overlaid as black crosses.}
\label{Fig10}
\end{figure*}

Methanol is mostly released into the gas phase through chemical desorption upon two-body surface reactions \citep{2016A&A...591A..52E}. Although purely gas-phase formation routes for CH$_3$OH have been proposed, they cannot reproduce the observed fractional abundances of $\geq10^{-9}$ \citep[e.g.,][]{2006FaDi..133...51G}. More recently, \citet{2017AJ....154...38H} showed that non-dissociative C-shocks maintain high CH$_3$OH abundances over a broad range of shock velocities including $v_\mathrm{shock}>40$\kms. Once in the gas phase, CH$_3$OH is rapidly destroyed on timescales of $10^{4}$–$10^{5}$ yr, especially under high cosmic-ray ionisation rates, so its widespread detection across the CMZ implies a continuous replenishment mechanism \citep{2013ApJ...764L..19Y}. In our target region, located at the interface where the dust-lane gas overshoots the CMZ, such replenishment is naturally expected: hydrodynamical simulations show that only $\sim30\%$ of the inflowing gas is accreted onto the CMZ, while the remainder overshoots, producing strong shocks with a net inflow rate of $\sim0.8\pm0.6$~M$_\odot$~yr$^{-1}$ \citep{2021ApJ...922...79H}. These large-scale shocks thus provide a steady supply of freshly liberated CH$_3$OH and SiO, maintaining the high abundances we observe.

\subsection{Association with star formation activity?}

Class~I masers are often associated with protostellar outflows \citep[e.g.,][]{2012ApJ...760L..20C}. To investigate whether the large scale maser emission is associated with protostellar activity, we use the Herschel Hi-GAL compact source catalogue \citep{{2017MNRAS.471..100E},{2021MNRAS.504.2742E}}, which provides a band-merged list of cold dust clumps identified across the 70--500~$\mu$m Herschel bands. In this catalogue, distances are derived following the methodology of \citet{2021A&A...646A..74M}, and physical parameters are obtained from single-temperature modified blackbody fits to the spectral energy distributions. The catalogue classifies clumps into three categories: starless (unbound and lacking signs of gravitational collapse), prestellar (gravitationally bound but not associated with 70~$\mu$m emission), and protostellar (showing evidence of active star formation, typically indicated by a 70~$\mu$m detection).

Using the Hi-GAL catalogue, we identified eight dust clumps in the region, five with reliable velocity estimates, and three without associated velocity information. Among these, five are classified as prestellar and three as protostellar. The three prestellar clumps lacking distance estimates in the Hi-GAL catalogue do not spatially coincide with the maser emission (marked as circled crosses in Fig.~\ref{Fig8}) and are therefore excluded from further analysis. For the remaining five clumps, the line-of-sight velocities are consistent with the CH$_3$OH emission (within 12\kms\ of the systemic LSR velocity of 105~\kms). Properties of these clumps are listed in Table~\ref{TableA1}, and their peak positions are indicated in Fig.~\ref{Fig8}. None of the protostellar clumps are located within 1~pc from the maser peaks, except for HIGALBM1.3390+0.2649, which lies approximately 1~pc in projection from the second brightest maser, G1.341+0.258.

Notably, four of the five clumps lie close to peaks in SiO emission. However, we find no Hi-GAL clumps associated with maser features in the southern and south-eastern region. The nearest clumps in these directions are over 7~pc away. While Class~I masers are often offset by 0.1–1.0~pc from Class~II masers, ultracompact HII regions, and mid-infrared sources \citep{2010A&A...517A..56F}, the large offsets of southern and south-eastern masers suggest that they are unlikely to be associated with the identified protostellar clumps. In contrast, the more localised maser emission near the thermal CH$_3$OH and SiO emission peaks may be physically linked to ongoing star formation activity traced by the nearby Hi-GAL clumps. No young stellar objects, extended green objects (EGOs), HII regions, or supernova remnants were found in the literature within a 1.5$'$ ($\sim$3.6~pc) radius of the southern and south-eastern maser peaks, suggesting a lack of known star formation or feedback activity.

\section{Discussion}

Class~I methanol masers are known to be associated with cloud–cloud collisions \citep{{1992SvA....36..590S}}. Given that our region of interest lies above the G1.3 cloud, where the near-side bar-driven inflow impacts the edge of the CMZ, overshoots, and continues as the high-velocity Helix stream, the observed parsec-scale maser emission may originate from the extremely dynamical environment of this region. To investigate the molecular gas kinematics underlying this emission, we use the $^{12}$CO (3–2) data from the CHIMPS2 survey. The integrated intensity map of the high velocity emission spanning 100 to 200\kms\ towards the CMZ region is presented in Fig.~\ref{Fig1}. In the integrated intensity map, we identify several emission features. There are elongated filamentary emission features  identified toward the SgrB region, one of the most massive star forming regions of our Galaxy \citep{{2016A&A...588A.143S},{2017A&A...604A...6S},{2018ApJ...853..171G},{2019A&A...628A...6S}}. At around $l\sim1.3^\circ$ we identify bright emission corresponding to the G1.3 cloud. The high velocity emission further arches upward forming another bright clumpy feature above G1.3 around $b\sim0.25^\circ$, which is the region of interest in our study. The emission further extends upward and then forms the Helix stream toward lower Galactic longitudes. Overall the brightest high velocity emission is detected towards the edge of the CMZ, around $l\sim1.2^\circ-1.5^\circ$. 

To investigate the origin of the maser emission toward the bright clump above the G1.3 cloud, we perform a kinematic analysis using the PV diagram. We extract a PV cut along line AB shown in Fig.~\ref{Fig1}, and the resulting diagram is presented in Fig.~\ref{Fig10} (left). The chosen cut spans $\sim115$~pc, intersecting the Galactic plane, the edge of the CMZ, and the high velocity Helix stream. The PV diagram (Fig.~\ref{Fig10}, left) reveals a broad velocity feature centred at $\sim$100\kms, along with high-velocity extensions reaching 150–200\kms. A striking velocity spike, spanning $\gtrsim120$~\kms\ near offsets 0.6$^\circ$–0.8$^\circ$, coincides with our region of interest. This feature is characteristic of an extended velocity feature \citep[EVF;][]{2019MNRAS.488.4663S}, a class of compact clouds with extreme velocity dispersions ($\Delta V > 100$~\kms), interpreted as arising from collisions between the CMZ and gas inflowing along bar dust lanes. \citet{2019MNRAS.488.4663S} identified an EVF at $l = 1.3^\circ$ and suggested that such regions should exhibit enhanced shock tracers and chemically rich gas. 

The spatial and kinematic continuity between G1.3, the EVF-like velocity feature, and the Helix stream places our target region within the proposed bar-inflow/CMZ-interaction scenario. Although the detailed nature of the large-scale gas streams in the Galactic centre remains an active topic of investigation, the extreme CO kinematics observed in the PV diagram indicate interacting molecular components with large velocity gradients. The widespread SiO emission further suggests that these interactions are accompanied by shocks. Examination of the PPV cube shows that the Class~I CH$_3$OH masers closely trace the high-velocity CO features, particularly near the EVF-like structure (Fig.~\ref{Fig10}, right). In this context, the parsec-scale Class~I CH$_3$OH masers are likely to trace shocks associated with the interacting gas components. At the same time, the presence of protostellar clumps near some thermally dominated CH$_3$OH peaks suggests that a subset of the masers may instead be linked to shocks driven by local star-formation activity.

\begin{figure}[!htbp]
\centering
\includegraphics[scale=0.45]{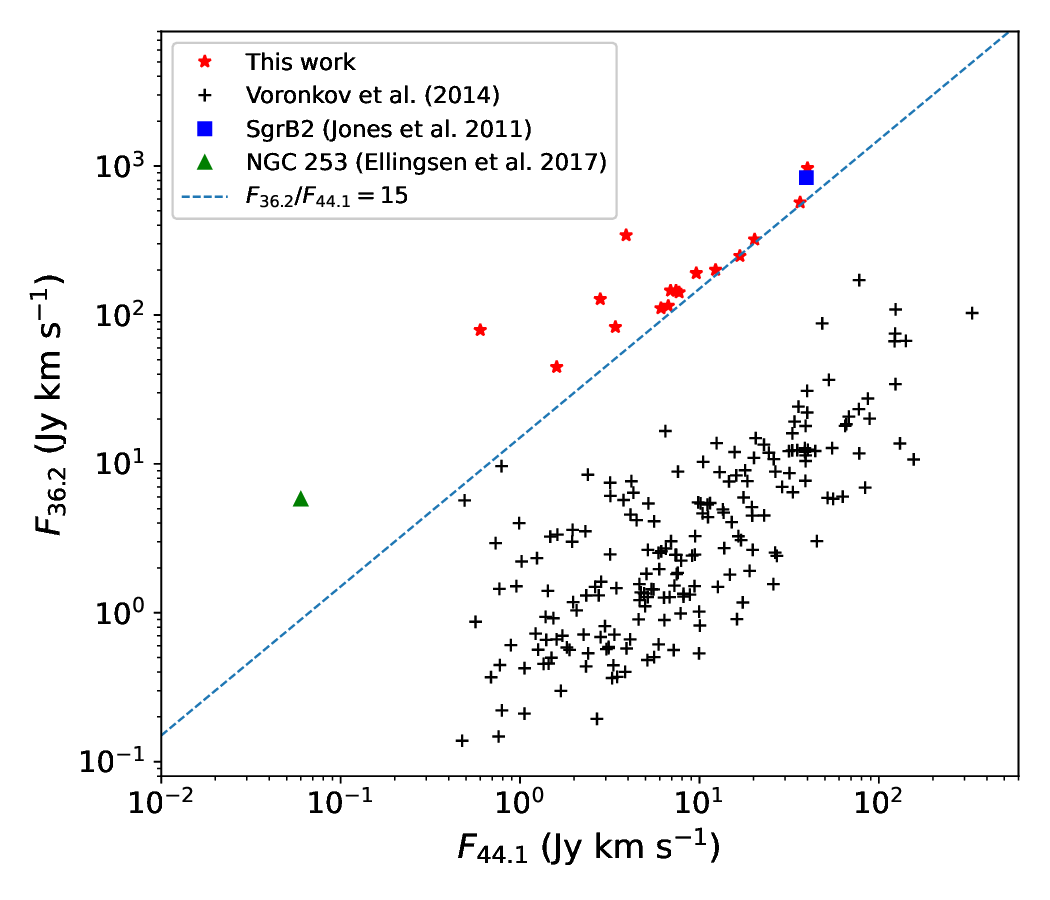}
\caption{Comparison of 36.2 and 44.1~GHz fluxes for Class~I methanol sources. Black crosses show Galactic masers from \citet{2014MNRAS.439.2584V}, red stars mark maser sources from this work, the blue square shows Sgr B2 in the Galactic centre from \citet{2011MNRAS.411.2293J}, and the green triangle denotes extragalactic source NGC 253 from \citet{2017MNRAS.472..604E}. The dashed line indicates a constant flux ratio $F_{36.2}/F_{44.1}$=15. For the sources in this work, the plotted values are fluxes integrated over the identical velocity ranges and may include both narrow maser emission and an underlying broader quasi-thermal component, whereas the literature values were derived using a variety of methods, including fitted emission components and peak-based estimates. The figure should therefore be interpreted as a qualitative comparison of the relative 36.2- and 44.1~GHz fluxes rather than as a strictly uniform maser-to-maser comparison.}
\label{Fig11}
\end{figure}

As shown in Section~3.2 and Fig.~\ref{Fig11}, our sources exhibit systematically enhanced observed 36.2/44.1 flux ratios compared to the bulk Galactic Class~I maser population \citep{2014MNRAS.439.2584V}. The excitation analysis of the 36.2, 44.1, and 95~GHz Class~I CH$_3$OH maser transitions by \citet{2016A&A...592A..31L} show that the 44.1~GHz transition is brighter than the 36.2~GHz line over much of the explored parameter space, while enhanced 36.2/44.1 ratios occur only in a more restricted regime. In their models, this regime corresponds to densities $\sim$$10^{6}$--$10^{7}$~cm$^{-3}$ and kinetic temperatures of $\sim$40--120 K, rather than the high temperature regime (200--400 K). They further argued that such conditions are more likely to arise in regions with little or no active star formation. To check whether the adopted excitation temperature is reasonable, we performed simple RADEX calculations for the thermal CH$_3$OH transition at 48.4 GHz under representative dense-gas conditions ($n(\mathrm H_2):10^6-5\times10^6$~cm$^{-3}$, $\Delta\mathrm V:30-40$\kms, $N_\mathrm{CH_3OH}=1-3\times10^{16}$~cm$^{-2}$). Resulting models give $T_\mathrm{ex}\sim$25--32 K, close to the fiducial value of 20~K assumed in the LTE estimate. 

The observed 36.2/44.1 flux ratio in our region is similar to observed Galactic-centre and extragalactic environments. Toward Sgr~B2, \citet{2011MNRAS.411.2293J} report 36.2 and 44.1~GHz methanol emission; from their reported peak flux densities and line widths, we infer a 36.2/44.1 flux ratio of $\sim$21, comparable to the values measured here. However, Sgr~B2 is a much more complex environment, with active high-mass star formation, multiple \hii ~regions, and both maser and quasi-thermal methanol emission \citep{{1997ApJ...474..346M},{2011MNRAS.411.2293J},{2022A&A...666A..31M}}. A more direct Galactic-centre analogue may be G+0.693$-$0.03, where the 36~GHz line is more than an order of magnitude brighter than the 44~GHz transition and the methanol emission has been interpreted as arising from large-scale shocks associated with cloud--cloud collisions rather than protostellar outflows \citep{2020MNRAS.497.4896Z}. Likewise, the CMZ cloud G1.6$-$0.025, near the eastern edge of the CMZ, shows 36.2~GHz maser emission superposed on a broader thermal component, evidence for cloud--cloud interactions, and little sign of ongoing massive star formation \citep{2009ApJ...692...47M}. This source has in turn been argued to resemble the extragalactic Class~I maser regions in NGC~253 and NGC~4945, where the 44.1~GHz transition is much weaker than the accompanying 36.2~GHz emission \citep{2018MNRAS.480.4578M}. 

In NGC~253, the 36.2~GHz masers are located near the interface between the nuclear ring and the bar ends, where large-scale cloud--cloud collisions are expected, while the 44.1~GHz emission is about two orders of magnitude weaker than the 36.2~GHz line \citep{2017MNRAS.472..604E}. Taken together, these comparisons suggest that the methanol emission in our bar-driven inflow region is akin to shock-dominated Class~I CH$_3$OH sources associated with extreme nuclear gas dynamics than to conventional Galactic star-formation outflows. We stress, however, that this comparison is qualitative. Our fluxes were measured by integrating over selected velocity ranges and may therefore include a broad quasi-thermal contribution, whereas literature values are often derived from fitted emission components, which in some cases may themselves include broader or quasi-thermal contributions. 

In the barred spiral galaxy M83, \citet{2019ApJ...884..100H} reported enhanced abundances of shock and dense-core tracers in one of the orbit–intersection regions, whereas the other showed a similar enhancement of shock tracers but little variation in dense-gas tracers and no evidence of star formation. This was interpreted as signatures of collision-induced cloud evolution that may, or may not, proceed to subsequent star formation. The region studied here provides a Galactic analogue in which the chemical and dynamical evolution of colliding gas flows can be probed down to sub-parsec scales, together with the onset of star formation under extreme cloud-collision conditions \citep{2015MNRAS.453.2471B,2020ApJ...891..168W}. This interpretation is further supported by recent extragalactic class~I CH$_3$OH maser detections in bar-driven environments. In particular, the 36.2~GHz maser in NGC~1365 is localized to the southern bar-inflow lane, where no prominent star formation is observed, and has been interpreted as tracing low-velocity, non-dissociative shocks rather than stellar feedback \citep{2026ApJ...999L..20C}. Together with the analogous case of NGC~253, this supports the idea that class~I methanol masers can selectively trace large-scale shock interfaces produced by bar-driven inflow. Our results therefore highlight the uniqueness of this CMZ boundary region, where inflow, feedback, and star formation processes converge. To disentangle the respective roles of large-scale shocks and embedded protostars in powering the observed maser emission, high-angular-resolution molecular-line and continuum observations will be essential to spatially resolve the individual maser components and obtain more reliable flux measurements.

\section{Conclusions}

We carried out the first dedicated search for Class~I methanol masers toward the transitional region between the G1.3 cloud and the Helix stream, where gas inflowing along the near-side dust lane has been proposed to interact with the CMZ boundary and arch upward. We detect widespread Class~I CH$_3$OH maser emission at 36.2~GHz and two candidate masers at 44.1~GHz. The brightest 36.2 GHz maser has an isotropic luminosity of $0.9\times10^{-3}$~L$_\odot$, placing it among the most luminous Class~I CH$_3$OH masers known in the Galaxy. All identified masers are part of a large maser complex extending over several parsecs. We also detect thermal CH$_3$OH emission at 48.4~GHz and morphologically correlated SiO emission over $\sim$24~pc. Both CH$_3$OH ($\sim$10$^{-6}$) and SiO ($\sim$10$^{-8}$) fractional abundances are strongly enhanced relative to those in typical spiral-arm clouds, and show a positive correlation with each other. We therefore conclude that the observed maser emission primarily traces shock-processed gas in the bar--CMZ interface region, where large-scale, turbulent gas interactions associated with dust-lane inflow are likely to play an important role. This interpretation is supported by the systematically enhanced 36.2/44.1 line ratios, lack of clear protostellar outflow tracers, and the close association of the maser velocities with the extended high-velocity ($\sim$100\kms) CO emission feature. Taken together, these properties make this region a promising Galactic analogue of shock-dominated class~I CH$_3$OH masers environments observed in nuclear regions of external barred galaxies, while any contribution from ongoing star formation appears to be secondary.

\begin{acknowledgements}

We thank the referee for the valuable comments and suggestions that improved the quality of the paper. We gratefully acknowledge late Prof. Dr. Karl Martin Menten for his guidance and support. We thank the staff of Yebes 40m telescope for their observational assistance. The 40 m radio telescope at Yebes Observatory is operated by the Spanish Geographic Institute (IGN; Ministerio de Transportes y Movilidad Sostenible).
V.V.S acknowledges the support of the Department of Atomic Energy, Government of India, under Project Identification No. RTI 4012. A.S-M.\ acknowledges support from the PID2023-146675NB grant funded by MCIN/AEI/10.13039/501100011033, and by the Spanish program Unidad de Excelencia María de Maeztu CEX2020-001058-M, financed by MCIN/AEI/10.13039/501100011033, and by the MaX-CSIC Excellence Award MaX4-SOMMA-ICE. G.E. acknowledges support from the Spanish grant PID2022-137980NB-I00, funded by MCIN/AEI/10.13039/501100011033/FEDER UE. M.C.S acknowledges financial support from the European Research Council under the ERC Starting Grant "GalFlow" (grant 101116226) and from the Fondazione Cariplo under the grant ERC attrattivit\`{a} n. 2023-3014.

\end{acknowledgements} 

\bibliographystyle{aa}
\bibliography{astro}

\begin{thebibliography}{64}
\expandafter\ifx\csname natexlab\endcsname\relax\def\natexlab#1{#1}\fi

\bibitem[{{Bae} {et~al.}(2011){Bae}, {Kim}, {Youn}, {Kim}, {Byun}, {Kang}, \&
  {Oh}}]{2011ApJS..196...21B}
{Bae}, J.-H., {Kim}, K.-T., {Youn}, S.-Y., {et~al.} 2011, \apjs, 196, 21

\bibitem[{{Balfour} {et~al.}(2015){Balfour}, {Whitworth}, {Hubber}, \&
  {Jaffa}}]{2015MNRAS.453.2471B}
{Balfour}, S.~K., {Whitworth}, A.~P., {Hubber}, D.~A., \& {Jaffa}, S.~E. 2015,
  \mnras, 453, 2471

\bibitem[{{Binney} {et~al.}(1991){Binney}, {Gerhard}, {Stark}, {Bally}, \&
  {Uchida}}]{1991MNRAS.252..210B}
{Binney}, J., {Gerhard}, O.~E., {Stark}, A.~A., {Bally}, J., \& {Uchida}, K.~I.
  1991, \mnras, 252, 210

\bibitem[{{Busch} {et~al.}(2022){Busch}, {Riquelme}, {G{\"u}sten}, {Menten},
  {Pillai}, \& {Kauffmann}}]{2022A&A...668A.183B}
{Busch}, L.~A., {Riquelme}, D., {G{\"u}sten}, R., {et~al.} 2022, \aap, 668,
  A183

\bibitem[{{Butterfield} {et~al.}(2022){Butterfield}, {Lang}, {Ginsburg},
  {Morris}, {Ott}, \& {Ludovici}}]{2022ApJ...936..186B}
{Butterfield}, N.~O., {Lang}, C.~C., {Ginsburg}, A., {et~al.} 2022, \apj, 936,
  186

\bibitem[{{Caswell} {et~al.}(2010){Caswell}, {Fuller}, {Green}, {Avison},
  {Breen}, {Brooks}, {Burton}, {Chrysostomou}, {Cox}, {Diamond}, {Ellingsen},
  {Gray}, {Hoare}, {Masheder}, {McClure-Griffiths}, {Pestalozzi}, {Phillips},
  {Quinn}, {Thompson}, {Voronkov}, {Walsh}, {Ward-Thompson}, {Wong-McSweeney},
  {Yates}, \& {Cohen}}]{2010MNRAS.404.1029C}
{Caswell}, J.~L., {Fuller}, G.~A., {Green}, J.~A., {et~al.} 2010, \mnras, 404,
  1029

\bibitem[{{Chen} {et~al.}(2026){Chen}, {Yang}, {Song}, {Ellingsen}, {Breen},
  {McCarthy}, {Yang}, \& {Mao}}]{2026ApJ...999L..20C}
{Chen}, X., {Yang}, T., {Song}, S., {et~al.} 2026, \apjl, 999, L20

\bibitem[{{Cotton} \& {Yusef-Zadeh}(2016)}]{2016ApJS..227...10C}
{Cotton}, W.~D. \& {Yusef-Zadeh}, F. 2016, \apjs, 227, 10

\bibitem[{{Cyganowski} {et~al.}(2012){Cyganowski}, {Brogan}, {Hunter}, {Zhang},
  {Friesen}, {Indebetouw}, \& {Chandler}}]{2012ApJ...760L..20C}
{Cyganowski}, C.~J., {Brogan}, C.~L., {Hunter}, T.~R., {et~al.} 2012, \apjl,
  760, L20

\bibitem[{{Eden} {et~al.}(2020){Eden}, {Moore}, {Currie}, {Rigby},
  {Rosolowsky}, {Su}, {Kim}, {Parsons}, {Morata}, {Chen}, {Minamidani}, {Park},
  {Ragan}, {Urquhart}, {Rani}, {Tahani}, {Billington}, {Deb}, {Figura},
  {Fujiyoshi}, {Joncas}, {Liao}, {Liu}, {Ma}, {Tuan-Anh}, {Yun}, {Zhang},
  {Zhu}, {Henshaw}, {Longmore}, {Kobayashi}, {Thompson}, {Ao},
  {Campbell-White}, {Ching}, {Chung}, {Duarte-Cabral}, {Fich}, {Gao}, {Graves},
  {Jiang}, {Kemper}, {Kuan}, {Kwon}, {Lee}, {Lee}, {Liu}, {Pe{\~n}aloza},
  {Peretto}, {Phuong}, {Pineda}, {Plume}, {Puspitaningrum}, {Samal}, {Soam},
  {Sun}, {Tang}, {Traficante}, {White}, {Yan}, {Yang}, {Yuan}, {Yue}, {Bemis},
  {Brunt}, {Chen}, {Cho}, {Clark}, {Cyganowski}, {Friberg}, {Fuller}, {Han},
  {Hoare}, {Izumi}, {Kim}, {Kim}, {Kim}, {Koch}, {Kuno}, {Lacialle}, {Lai},
  {Lee}, {Lee}, {Li}, {Liu}, {Mairs}, {Pan}, {Qian}, {Scicluna}, {Shi}, {Shi},
  {Srinivasan}, {Tan}, {Thomas}, {Torii}, {Trejo}, {Umemoto}, {Violino},
  {Wallstr{\"o}m}, {Wang}, {Wu}, {Yuan}, {Zhang}, {Zhang}, {Zhou}, \&
  {Zhou}}]{2020MNRAS.498.5936E}
{Eden}, D.~J., {Moore}, T.~J.~T., {Currie}, M.~J., {et~al.} 2020, \mnras, 498,
  5936

\bibitem[{{Elia} {et~al.}(2021){Elia}, {Merello}, {Molinari}, {Schisano},
  {Zavagno}, {Russeil}, {M{\`e}ge}, {Martin}, {Olmi}, {Pestalozzi}, {Plume},
  {Ragan}, {Benedettini}, {Eden}, {Moore}, {Noriega-Crespo}, {Paladini},
  {Palmeirim}, {Pezzuto}, {Pilbratt}, {Rygl}, {Schilke}, {Strafella}, {Tan},
  {Traficante}, {Baldeschi}, {Bally}, {di Giorgio}, {Fiorellino}, {Liu},
  {Piazzo}, \& {Polychroni}}]{2021MNRAS.504.2742E}
{Elia}, D., {Merello}, M., {Molinari}, S., {et~al.} 2021, \mnras, 504, 2742

\bibitem[{{Elia} {et~al.}(2017){Elia}, {Molinari}, {Schisano}, {Pestalozzi},
  {Pezzuto}, {Merello}, {Noriega-Crespo}, {Moore}, {Russeil}, {Mottram},
  {Paladini}, {Strafella}, {Benedettini}, {Bernard}, {Di Giorgio}, {Eden},
  {Fukui}, {Plume}, {Bally}, {Martin}, {Ragan}, {Jaffa}, {Motte}, {Olmi},
  {Schneider}, {Testi}, {Wyrowski}, {Zavagno}, {Calzoletti}, {Faustini},
  {Natoli}, {Palmeirim}, {Piacentini}, {Piazzo}, {Pilbratt}, {Polychroni},
  {Baldeschi}, {Beltr{\'a}n}, {Billot}, {Cambr{\'e}sy}, {Cesaroni},
  {Garc{\'\i}a-Lario}, {Hoare}, {Huang}, {Joncas}, {Liu}, {Maiolo}, {Marsh},
  {Maruccia}, {M{\`e}ge}, {Peretto}, {Rygl}, {Schilke}, {Thompson},
  {Traficante}, {Umana}, {Veneziani}, {Ward-Thompson}, {Whitworth}, {Arab},
  {Bandieramonte}, {Becciani}, {Brescia}, {Buemi}, {Bufano}, {Butora},
  {Cavuoti}, {Costa}, {Fiorellino}, {Hajnal}, {Hayakawa}, {Kacsuk}, {Leto}, {Li
  Causi}, {Marchili}, {Martinavarro-Armengol}, {Mercurio}, {Molinaro},
  {Riccio}, {Sano}, {Sciacca}, {Tachihara}, {Torii}, {Trigilio}, {Vitello}, \&
  {Yamamoto}}]{2017MNRAS.471..100E}
{Elia}, D., {Molinari}, S., {Schisano}, E., {et~al.} 2017, \mnras, 471, 100

\bibitem[{{Ellingsen} {et~al.}(2017){Ellingsen}, {Chen}, {Breen}, \&
  {Qiao}}]{2017MNRAS.472..604E}
{Ellingsen}, S.~P., {Chen}, X., {Breen}, S.~L., \& {Qiao}, H.~H. 2017, \mnras,
  472, 604

\bibitem[{{Ellingsen} {et~al.}(2014){Ellingsen}, {Chen}, {Qiao}, {Baan}, {An},
  {Li}, \& {Breen}}]{2014ApJ...790L..28E}
{Ellingsen}, S.~P., {Chen}, X., {Qiao}, H.-H., {et~al.} 2014, \apjl, 790, L28

\bibitem[{{Esplugues} {et~al.}(2016){Esplugues}, {Cazaux}, {Meijerink},
  {Spaans}, \& {Caselli}}]{2016A&A...591A..52E}
{Esplugues}, G.~B., {Cazaux}, S., {Meijerink}, R., {Spaans}, M., \& {Caselli},
  P. 2016, \aap, 591, A52

\bibitem[{{Fontani} {et~al.}(2010){Fontani}, {Cesaroni}, \&
  {Furuya}}]{2010A&A...517A..56F}
{Fontani}, F., {Cesaroni}, R., \& {Furuya}, R.~S. 2010, \aap, 517, A56

\bibitem[{{Garrod} {et~al.}(2006){Garrod}, {Park}, {Caselli}, \&
  {Herbst}}]{2006FaDi..133...51G}
{Garrod}, R., {Park}, I.~H., {Caselli}, P., \& {Herbst}, E. 2006, Faraday
  Discussions, 133, 51

\bibitem[{{Ghez} {et~al.}(2005){Ghez}, {Salim}, {Hornstein}, {Tanner}, {Lu},
  {Morris}, {Becklin}, \& {Duch{\^e}ne}}]{2005ApJ...620..744G}
{Ghez}, A.~M., {Salim}, S., {Hornstein}, S.~D., {et~al.} 2005, \apj, 620, 744

\bibitem[{{Ginsburg} {et~al.}(2018){Ginsburg}, {Bally}, {Barnes}, {Bastian},
  {Battersby}, {Beuther}, {Brogan}, {Contreras}, {Corby}, {Darling}, {De Pree},
  {Galv{\'a}n-Madrid}, {Garay}, {Henshaw}, {Hunter}, {Kruijssen}, {Longmore},
  {Lu}, {Meng}, {Mills}, {Ott}, {Pineda}, {S{\'a}nchez-Monge}, {Schilke},
  {Schmiedeke}, {Walker}, \& {Wilner}}]{2018ApJ...853..171G}
{Ginsburg}, A., {Bally}, J., {Barnes}, A., {et~al.} 2018, \apj, 853, 171

\bibitem[{{Ginsburg} {et~al.}(2016){Ginsburg}, {Henkel}, {Ao}, {Riquelme},
  {Kauffmann}, {Pillai}, {Mills}, {Requena-Torres}, {Immer}, {Testi}, {Ott},
  {Bally}, {Battersby}, {Darling}, {Aalto}, {Stanke}, {Kendrew}, {Kruijssen},
  {Longmore}, {Dale}, {Guesten}, \& {Menten}}]{2016A&A...586A..50G}
{Ginsburg}, A., {Henkel}, C., {Ao}, Y., {et~al.} 2016, \aap, 586, A50

\bibitem[{{GRAVITY Collaboration} {et~al.}(2019){GRAVITY Collaboration},
  {Abuter}, {Amorim}, {Baub{\"o}ck}, {Berger}, {Bonnet}, {Brandner},
  {Cl{\'e}net}, {Coud{\'e} Du Foresto}, {de Zeeuw}, {Dexter}, {Duvert},
  {Eckart}, {Eisenhauer}, {F{\"o}rster Schreiber}, {Garcia}, {Gao}, {Gendron},
  {Genzel}, {Gerhard}, {Gillessen}, {Habibi}, {Haubois}, {Henning}, {Hippler},
  {Horrobin}, {Jim{\'e}nez-Rosales}, {Jocou}, {Kervella}, {Lacour},
  {Lapeyr{\`e}re}, {Le Bouquin}, {L{\'e}na}, {Ott}, {Paumard}, {Perraut},
  {Perrin}, {Pfuhl}, {Rabien}, {Rodriguez Coira}, {Rousset}, {Scheithauer},
  {Sternberg}, {Straub}, {Straubmeier}, {Sturm}, {Tacconi}, {Vincent}, {von
  Fellenberg}, {Waisberg}, {Widmann}, {Wieprecht}, {Wiezorrek}, {Woillez}, \&
  {Yazici}}]{2019A&A...625L..10G}
{GRAVITY Collaboration}, {Abuter}, R., {Amorim}, A., {et~al.} 2019, \aap, 625,
  L10

\bibitem[{{Harada} {et~al.}(2019){Harada}, {Sakamoto}, {Mart{\'\i}n},
  {Watanabe}, {Aladro}, {Riquelme}, \& {Hirota}}]{2019ApJ...884..100H}
{Harada}, N., {Sakamoto}, K., {Mart{\'\i}n}, S., {et~al.} 2019, \apj, 884, 100

\bibitem[{{Hatchfield} {et~al.}(2021){Hatchfield}, {Sormani}, {Tress},
  {Battersby}, {Smith}, {Glover}, \& {Klessen}}]{2021ApJ...922...79H}
{Hatchfield}, H.~P., {Sormani}, M.~C., {Tress}, R.~G., {et~al.} 2021, \apj,
  922, 79

\bibitem[{{Henshaw} {et~al.}(2023){Henshaw}, {Barnes}, {Battersby}, {Ginsburg},
  {Sormani}, \& {Walker}}]{2023ASPC..534...83H}
{Henshaw}, J.~D., {Barnes}, A.~T., {Battersby}, C., {et~al.} 2023, in
  Astronomical Society of the Pacific Conference Series, Vol. 534, Protostars
  and Planets VII, ed. S.~{Inutsuka}, Y.~{Aikawa}, T.~{Muto}, K.~{Tomida}, \&
  M.~{Tamura}, 83

\bibitem[{{Holdship} {et~al.}(2017){Holdship}, {Viti}, {Jim{\'e}nez-Serra},
  {Makrymallis}, \& {Priestley}}]{2017AJ....154...38H}
{Holdship}, J., {Viti}, S., {Jim{\'e}nez-Serra}, I., {Makrymallis}, A., \&
  {Priestley}, F. 2017, \aj, 154, 38

\bibitem[{{Humire} {et~al.}(2020){Humire}, {Henkel}, {Gong}, {Leurini},
  {Mauersberger}, {Levshakov}, {Winkel}, {Tarchi}, {Castangia}, {Malawi},
  {Asiri}, {Ellingsen}, {McCarthy}, {Chen}, \& {Tang}}]{2020A&A...633A.106H}
{Humire}, P.~K., {Henkel}, C., {Gong}, Y., {et~al.} 2020, \aap, 633, A106

\bibitem[{{Humire} {et~al.}(2022){Humire}, {Henkel}, {Hern{\'a}ndez-G{\'o}mez},
  {Mart{\'\i}n}, {Mangum}, {Harada}, {Muller}, {Sakamoto}, {Tanaka},
  {Yoshimura}, {Nakanishi}, {M{\"u}hle}, {Herrero-Illana}, {Meier}, {Caux},
  {Aladro}, {Mauersberger}, {Viti}, {Colzi}, {Rivilla}, {Gorski}, {Menten},
  {Huang}, {Aalto}, {van der Werf}, \& {Emig}}]{2022A&A...663A..33H}
{Humire}, P.~K., {Henkel}, C., {Hern{\'a}ndez-G{\'o}mez}, A., {et~al.} 2022,
  \aap, 663, A33

\bibitem[{{Jones} {et~al.}(2011){Jones}, {Burton}, {Tothill}, \&
  {Cunningham}}]{2011MNRAS.411.2293J}
{Jones}, P.~A., {Burton}, M.~G., {Tothill}, N.~F.~H., \& {Cunningham}, M.~R.
  2011, \mnras, 411, 2293

\bibitem[{{Kim} {et~al.}(2019){Kim}, {Kim}, \& {Kim}}]{2019ApJS..244....2K}
{Kim}, W.-J., {Kim}, K.-T., \& {Kim}, K.-T. 2019, \apjs, 244, 2

\bibitem[{{Kim} {et~al.}(2023){Kim}, {Urquhart}, {Veena}, {Fuller}, {Schilke},
  \& {Kim}}]{2023A&A...679A.123K}
{Kim}, W.-J., {Urquhart}, J.~S., {Veena}, V.~S., {et~al.} 2023, \aap, 679, A123

\bibitem[{{Kurtz} {et~al.}(2004){Kurtz}, {Hofner}, \&
  {{\'A}lvarez}}]{2004ApJS..155..149K}
{Kurtz}, S., {Hofner}, P., \& {{\'A}lvarez}, C.~V. 2004, \apjs, 155, 149

\bibitem[{{Leurini} {et~al.}(2016){Leurini}, {Menten}, \&
  {Walmsley}}]{2016A&A...592A..31L}
{Leurini}, S., {Menten}, K.~M., \& {Walmsley}, C.~M. 2016, \aap, 592, A31

\bibitem[{{Lindner} {et~al.}(2015){Lindner}, {Vera-Ciro}, {Murray},
  {Stanimirovi{\'c}}, {Babler}, {Heiles}, {Hennebelle}, {Goss}, \&
  {Dickey}}]{2015AJ....149..138L}
{Lindner}, R.~R., {Vera-Ciro}, C., {Murray}, C.~E., {et~al.} 2015, \aj, 149,
  138

\bibitem[{{Marsh} {et~al.}(2017){Marsh}, {Whitworth}, {Lomax}, {Ragan},
  {Becciani}, {Cambr{\'e}sy}, {Di Giorgio}, {Eden}, {Elia}, {Kacsuk},
  {Molinari}, {Palmeirim}, {Pezzuto}, {Schneider}, {Sciacca}, \&
  {Vitello}}]{2017MNRAS.471.2730M}
{Marsh}, K.~A., {Whitworth}, A.~P., {Lomax}, O., {et~al.} 2017, \mnras, 471,
  2730

\bibitem[{{McCarthy} {et~al.}(2018){McCarthy}, {Ellingsen}, {Breen}, {Henkel},
  {Voronkov}, \& {Chen}}]{2018MNRAS.480.4578M}
{McCarthy}, T.~P., {Ellingsen}, S.~P., {Breen}, S.~L., {et~al.} 2018, \mnras,
  480, 4578

\bibitem[{{M{\`e}ge} {et~al.}(2021){M{\`e}ge}, {Russeil}, {Zavagno}, {Elia},
  {Molinari}, {Brunt}, {Butora}, {Cambresy}, {Di Giorgio}, {Fenouillet},
  {Fukui}, {Lambert}, {Makai}, {Merello}, {Meunier}, {Molinaro}, {Moreau},
  {Pezzuto}, {Poulin}, {Schisano}, \& {Schuller}}]{2021A&A...646A..74M}
{M{\`e}ge}, P., {Russeil}, D., {Zavagno}, A., {et~al.} 2021, \aap, 646, A74

\bibitem[{{Mehringer} \& {Menten}(1997)}]{1997ApJ...474..346M}
{Mehringer}, D.~M. \& {Menten}, K.~M. 1997, \apj, 474, 346

\bibitem[{{Meng} {et~al.}(2022){Meng}, {S{\'a}nchez-Monge}, {Schilke},
  {Ginsburg}, {DePree}, {Budaiev}, {Jeff}, {Schmiedeke}, {Schw{\"o}rer},
  {Veena}, \& {M{\"o}ller}}]{2022A&A...666A..31M}
{Meng}, F., {S{\'a}nchez-Monge}, {\'A}., {Schilke}, P., {et~al.} 2022, \aap,
  666, A31

\bibitem[{{Menten}(1991)}]{1991ASPC...16..119M}
{Menten}, K.~M. 1991, in Astronomical Society of the Pacific Conference Series,
  Vol.~16, Atoms, Ions and Molecules: New Results in Spectral Line
  Astrophysics, ed. A.~D. {Haschick} \& P.~T.~P. {Ho}, 119--136

\bibitem[{{Menten} {et~al.}(2009){Menten}, {Wilson}, {Leurini}, \&
  {Schilke}}]{2009ApJ...692...47M}
{Menten}, K.~M., {Wilson}, R.~W., {Leurini}, S., \& {Schilke}, P. 2009, \apj,
  692, 47

\bibitem[{{Minier} {et~al.}(2001){Minier}, {Conway}, \&
  {Booth}}]{2001A&A...369..278M}
{Minier}, V., {Conway}, J.~E., \& {Booth}, R.~S. 2001, \aap, 369, 278

\bibitem[{{Morris}(1993)}]{1993ApJ...408..496M}
{Morris}, M. 1993, \apj, 408, 496

\bibitem[{{Morris} \& {Serabyn}(1996)}]{1996ARA&A..34..645M}
{Morris}, M. \& {Serabyn}, E. 1996, \araa, 34, 645

\bibitem[{{M{\"u}ller} {et~al.}(2005){M{\"u}ller}, {Schl{\"o}der}, {Stutzki},
  \& {Winnewisser}}]{2005JMoSt.742..215M}
{M{\"u}ller}, H. S.~P., {Schl{\"o}der}, F., {Stutzki}, J., \& {Winnewisser}, G.
  2005, Journal of Molecular Structure, 742, 215

\bibitem[{{Ortiz-Le{\'o}n} {et~al.}(2021){Ortiz-Le{\'o}n}, {Menten},
  {Brunthaler}, {Csengeri}, {Urquhart}, {Wyrowski}, {Gong}, {Rugel}, {Dzib},
  {Yang}, {Nguyen}, {Cotton}, {Medina}, {Dokara}, {K{\"o}nig}, {Beuther},
  {Pandian}, {Reich}, \& {Roy}}]{2021A&A...651A..87O}
{Ortiz-Le{\'o}n}, G.~N., {Menten}, K.~M., {Brunthaler}, A., {et~al.} 2021,
  \aap, 651, A87

\bibitem[{{Pety}(2005)}]{pety2005_gildas}
{Pety}, J. 2005, in SF2A-2005: Semaine de l'Astrophysique Francaise, ed.
  F.~{Casoli}, T.~{Contini}, J.~M. {Hameury}, \& L.~{Pagani}, 721

\bibitem[{{Pihlstr{\"o}m} {et~al.}(2014){Pihlstr{\"o}m}, {Sjouwerman}, {Frail},
  {Claussen}, {Mesler}, \& {McEwen}}]{2014AJ....147...73P}
{Pihlstr{\"o}m}, Y.~M., {Sjouwerman}, L.~O., {Frail}, D.~A., {et~al.} 2014,
  \aj, 147, 73

\bibitem[{{Plambeck} \& {Menten}(1990)}]{1990ApJ...364..555P}
{Plambeck}, R.~L. \& {Menten}, K.~M. 1990, \apj, 364, 555

\bibitem[{{Pratap} {et~al.}(2008){Pratap}, {Shute}, {Keane}, {Battersby}, \&
  {Sterling}}]{2008AJ....135.1718P}
{Pratap}, P., {Shute}, P.~A., {Keane}, T.~C., {Battersby}, C., \& {Sterling},
  S. 2008, \aj, 135, 1718

\bibitem[{{Riener} {et~al.}(2019){Riener}, {Kainulainen}, {Henshaw}, {Orkisz},
  {Murray}, \& {Beuther}}]{2019A&A...628A..78R}
{Riener}, M., {Kainulainen}, J., {Henshaw}, J.~D., {et~al.} 2019, \aap, 628,
  A78

\bibitem[{{S{\'a}nchez-Monge} {et~al.}(2017){S{\'a}nchez-Monge}, {Schilke},
  {Schmiedeke}, {Ginsburg}, {Cesaroni}, {Lis}, {Qin}, {M{\"u}ller}, {Bergin},
  {Comito}, \& {M{\"o}ller}}]{2017A&A...604A...6S}
{S{\'a}nchez-Monge}, {\'A}., {Schilke}, P., {Schmiedeke}, A., {et~al.} 2017,
  \aap, 604, A6

\bibitem[{{Schilke} {et~al.}(1997){Schilke}, {Walmsley}, {Pineau des Forets},
  \& {Flower}}]{1997A&A...321..293S}
{Schilke}, P., {Walmsley}, C.~M., {Pineau des Forets}, G., \& {Flower}, D.~R.
  1997, \aap, 321, 293

\bibitem[{{Schmiedeke} {et~al.}(2016){Schmiedeke}, {Schilke}, {M{\"o}ller},
  {S{\'a}nchez-Monge}, {Bergin}, {Comito}, {Csengeri}, {Lis}, {Molinari},
  {Qin}, \& {Rolffs}}]{2016A&A...588A.143S}
{Schmiedeke}, A., {Schilke}, P., {M{\"o}ller}, T., {et~al.} 2016, \aap, 588,
  A143

\bibitem[{{Schw{\"o}rer} {et~al.}(2019){Schw{\"o}rer}, {S{\'a}nchez-Monge},
  {Schilke}, {M{\"o}ller}, {Ginsburg}, {Meng}, {Schmiedeke}, {M{\"u}ller},
  {Lis}, \& {Qin}}]{2019A&A...628A...6S}
{Schw{\"o}rer}, A., {S{\'a}nchez-Monge}, {\'A}., {Schilke}, P., {et~al.} 2019,
  \aap, 628, A6

\bibitem[{{Sobolev}(1992)}]{1992SvA....36..590S}
{Sobolev}, A.~M. 1992, \sovast, 36, 590

\bibitem[{{Sormani} {et~al.}(2019){Sormani}, {Tre{\ss}}, {Glover}, {Klessen},
  {Barnes}, {Battersby}, {Clark}, {Hatchfield}, \&
  {Smith}}]{2019MNRAS.488.4663S}
{Sormani}, M.~C., {Tre{\ss}}, R.~G., {Glover}, S. C.~O., {et~al.} 2019, \mnras,
  488, 4663

\bibitem[{{Tercero} {et~al.}(2021){Tercero}, {L{\'o}pez-P{\'e}rez}, {Gallego},
  {Beltr{\'a}n}, {Garc{\'\i}a}, {Patino-Esteban}, {L{\'o}pez-Fern{\'a}ndez},
  {G{\'o}mez-Molina}, {Diez}, {Garc{\'\i}a-Carre{\~n}o}, {Malo}, {Amils},
  {Serna}, {Albo}, {Hern{\'a}ndez}, {Vaquero}, {Gonz{\'a}lez-Garc{\'\i}a},
  {Barbas}, {L{\'o}pez-Fern{\'a}ndez}, {Bujarrabal}, {G{\'o}mez-Garrido},
  {Pardo}, {Santander-Garc{\'\i}a}, {Tercero}, {Cernicharo}, \& {de
  Vicente}}]{2021A&A...645A..37T}
{Tercero}, F., {L{\'o}pez-P{\'e}rez}, J.~A., {Gallego}, J.~D., {et~al.} 2021,
  \aap, 645, A37

\bibitem[{{Townes} {et~al.}(1983){Townes}, {Lacy}, {Geballe}, \&
  {Hollenbach}}]{1983Natur.301..661T}
{Townes}, C.~H., {Lacy}, J.~H., {Geballe}, T.~R., \& {Hollenbach}, D.~J. 1983,
  \nat, 301, 661

\bibitem[{{Veena} {et~al.}(2024){Veena}, {Kim}, {S{\'a}nchez-Monge}, {Schilke},
  {Menten}, {Fuller}, {Sormani}, {Wyrowski}, {Banda-Barrag{\'a}n}, {Riquelme},
  {Tarr{\'\i}o}, \& {de Vicente}}]{2024A&A...689A.121V}
{Veena}, V.~S., {Kim}, W.~J., {S{\'a}nchez-Monge}, {\'A}., {et~al.} 2024, \aap,
  689, A121

\bibitem[{{Voronkov} {et~al.}(2014){Voronkov}, {Caswell}, {Ellingsen}, {Green},
  \& {Breen}}]{2014MNRAS.439.2584V}
{Voronkov}, M.~A., {Caswell}, J.~L., {Ellingsen}, S.~P., {Green}, J.~A., \&
  {Breen}, S.~L. 2014, \mnras, 439, 2584

\bibitem[{{Wu} {et~al.}(2020){Wu}, {Tan}, {Christie}, \&
  {Nakamura}}]{2020ApJ...891..168W}
{Wu}, B., {Tan}, J.~C., {Christie}, D., \& {Nakamura}, F. 2020, \apj, 891, 168

\bibitem[{{Yang} {et~al.}(2023){Yang}, {Gong}, {Menten}, {Urquhart}, {Henkel},
  {Wyrowski}, {Csengeri}, {Ellingsen}, {Bemis}, \&
  {Jang}}]{2023A&A...675A.112Y}
{Yang}, W., {Gong}, Y., {Menten}, K.~M., {et~al.} 2023, \aap, 675, A112

\bibitem[{{Yusef-Zadeh} {et~al.}(2013){Yusef-Zadeh}, {Cotton}, {Viti},
  {Wardle}, \& {Royster}}]{2013ApJ...764L..19Y}
{Yusef-Zadeh}, F., {Cotton}, W., {Viti}, S., {Wardle}, M., \& {Royster}, M.
  2013, \apjl, 764, L19

\bibitem[{{Zeng} {et~al.}(2020){Zeng}, {Zhang}, {Jim{\'e}nez-Serra}, {Tercero},
  {Lu}, {Mart{\'\i}n-Pintado}, {de Vicente}, {Rivilla}, \&
  {Li}}]{2020MNRAS.497.4896Z}
{Zeng}, S., {Zhang}, Q., {Jim{\'e}nez-Serra}, I., {et~al.} 2020, \mnras, 497,
  4896

\end{thebibliography}

\clearpage
\newpage
\onecolumn
\appendix
\section{Additional figures and tables}
\begin{figure}[h]
\centering
\includegraphics[scale=0.5]{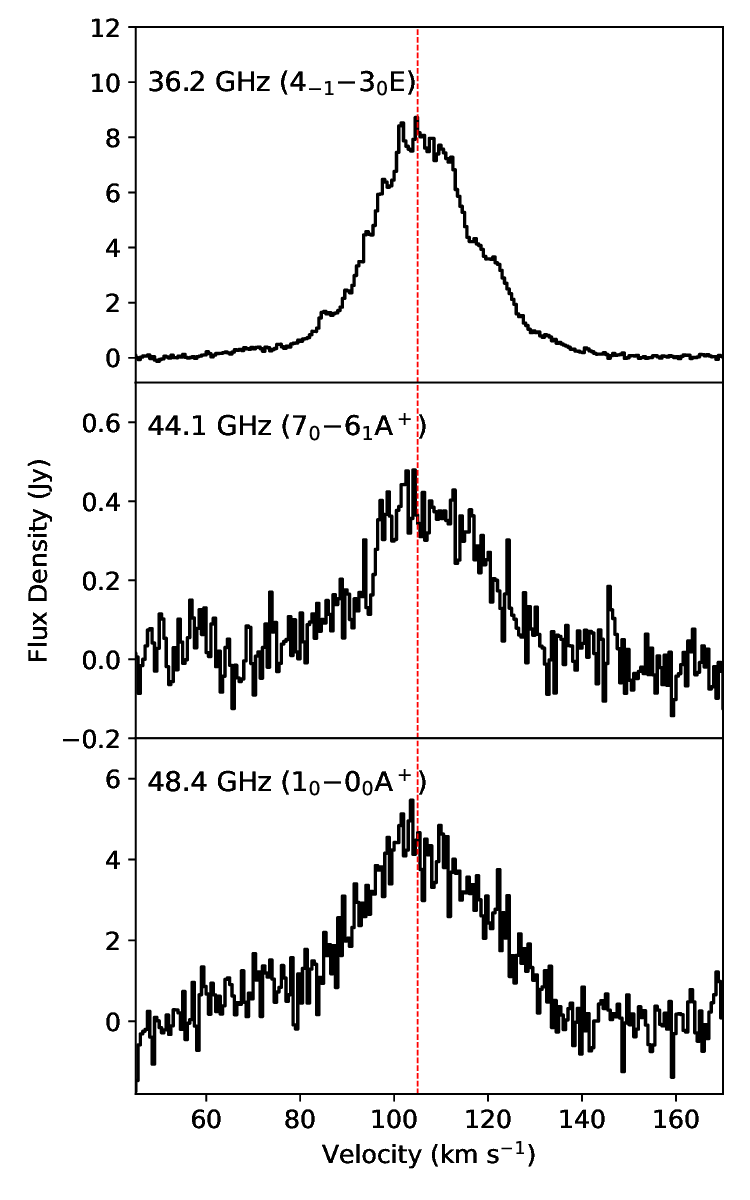}
\caption{Average spectra of CH$_3$OH emission at 36.2~GHz (top), 44.1~GHz (middle), and 48.4~GHz (bottom) extracted within a circular aperture of radius 205$''$. The spectra are obtained by summing the flux density over all pixels enclosed within the aperture. The vertical red dashed line marks LSR velocity of 105\kms.}
\label{FigA1}
\end{figure}


\begin{table*}[!htbp]
\centering
\footnotesize
\caption{Properties of Herschel Hi-GAL clumps with available velocity information}
\begin{tabular}{c c c c c c c c c c c}
\hline\\
Source ID&GLON&GLAT&Diameter&Distance$^*$&Velocity&Mass\tablefootmark{a} &$T_\textrm{dust}\tablefootmark{a}$&$L_\textrm{Bol}$&$\Sigma$&Class\tablefootmark{b}\\
&(Deg)&(Deg)&(pc)&(kpc)&(\kms)&(M$_\odot$)&(K)&(L$_\odot$)&(g\,cm$^{-2}$)&\\
\hline\\
HIGALBM1.2876+0.2928&1.2876&0.2928&1.5&8.6&110.6&3329&13.0&1750&0.4&2\\
HIGALBM1.2882+0.2743&1.2882&0.2743&1.0&8.6&103.1&5143&10.8&676&1.5&1\\
HIGALBM1.3052+0.2733&1.3052&0.2733&0.7&8.6&116.6&565&14.9&725&0.3&2\\
HIGALBM1.3390+0.2649&1.3390&0.2649&1.2&8.6&108.8&2195&13.0&1889&0.4&2\\
HIGALBM1.3435+0.2596&1.3435&0.2596&1.2&8.6&102.7&13739&10.3&1276&2.6&1\\
\hline
\end{tabular}
\tablefoot{\tablefoottext{*}{Heliocentric distance assigned in the Hi-GAL catalogue,} \tablefoottext{a}{Obtained by fitting the SEDs in the spectral range 160~$\mu$m$\leq\lambda\leq$500~$\mu$m} with a modified blackbody function, \tablefoottext{b}{1: prestellar, 2: protostellar}}
\label{TableA1}
\end{table*}

\begin{figure}[]
\centering
\includegraphics[scale=0.45]{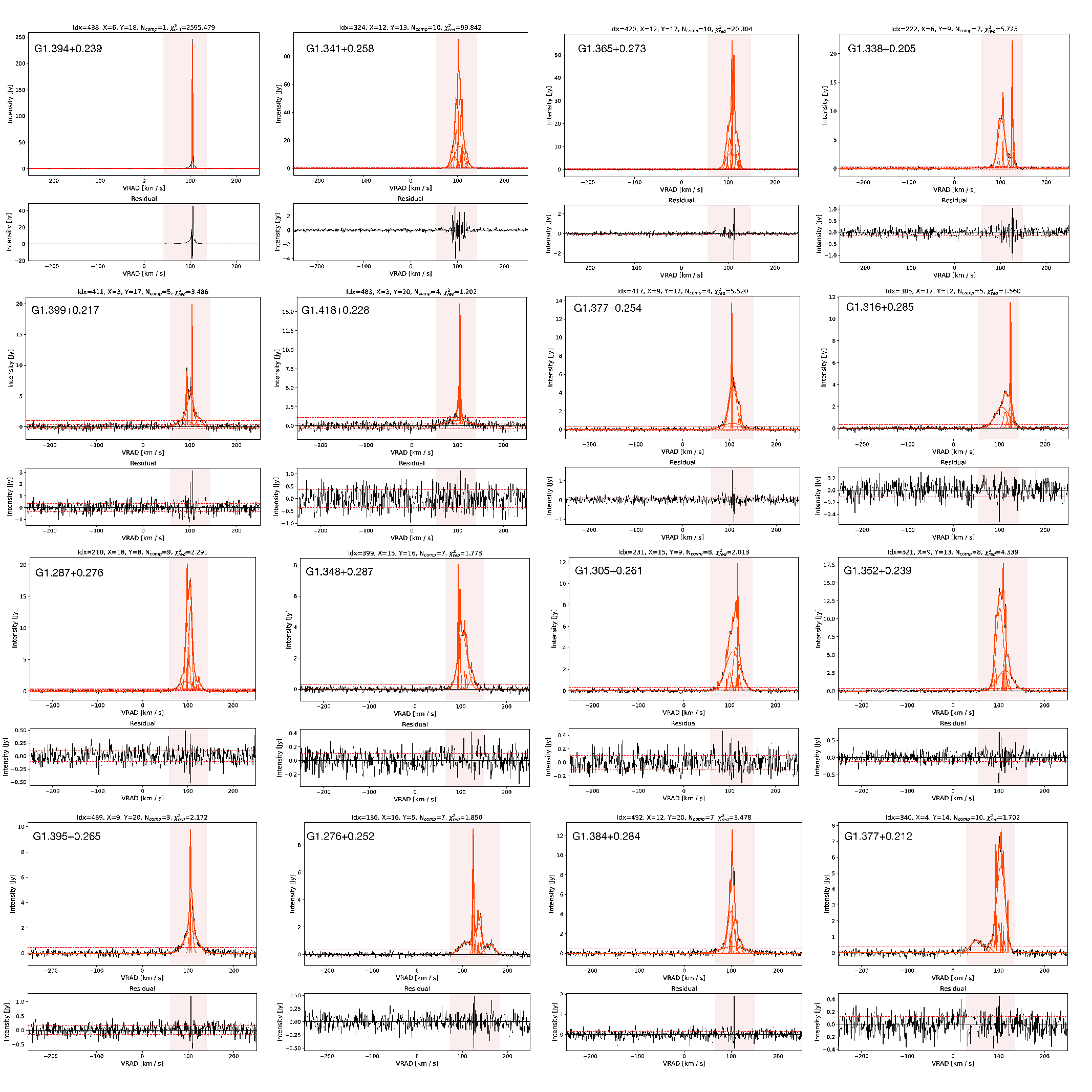}
\caption{GAUSSPY+ decomposition of the 36.2~GHz spectra towards the peak of 16 maser candidate positions. The top panels show the observed spectra (black), Gaussian model, and individual components (shaded red). The bottom panels show the residuals after model subtraction. Narrow, high-S/N components consistent with maser emission are clearly identifiable. }
\label{FigA2}
\end{figure}

\begin{figure}[]
\centering
\includegraphics[scale=0.3]{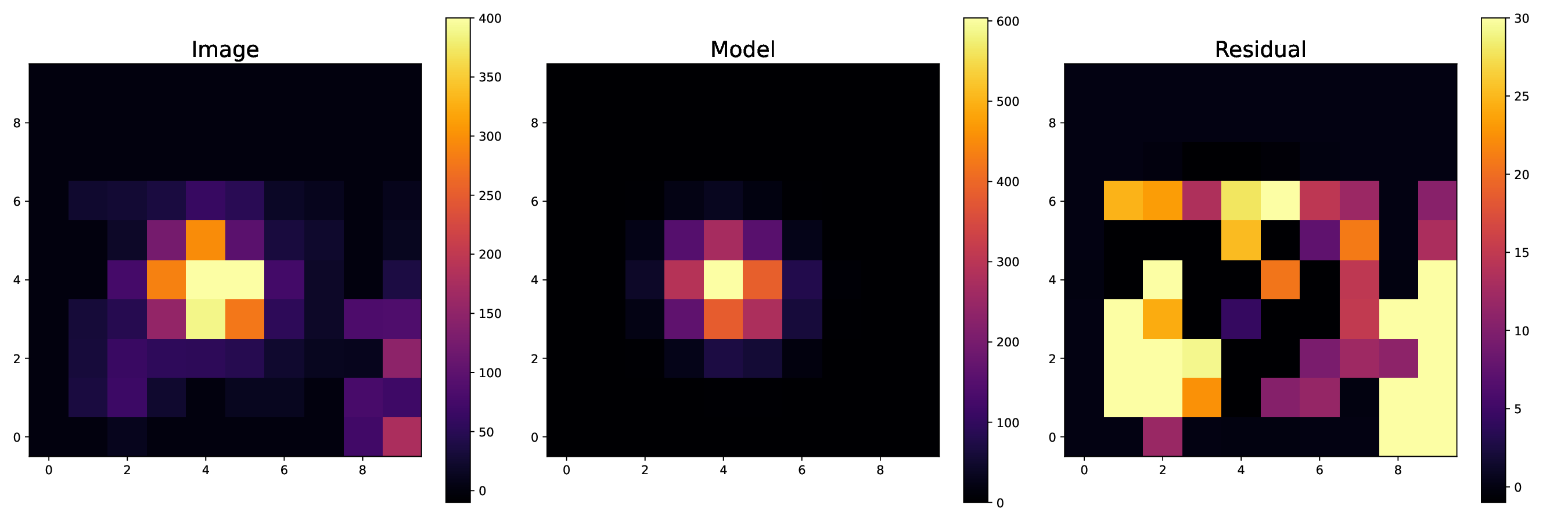}
\caption{Integrated intensity cutout of the brightest Class~I maser G1.3940+0.2390 (left), the best-fit two-dimensional Gaussian model (middle), and the residual map (right).}
\label{FigA3}
\end{figure}

\begin{figure}[]
\centering
\includegraphics[scale=0.52]{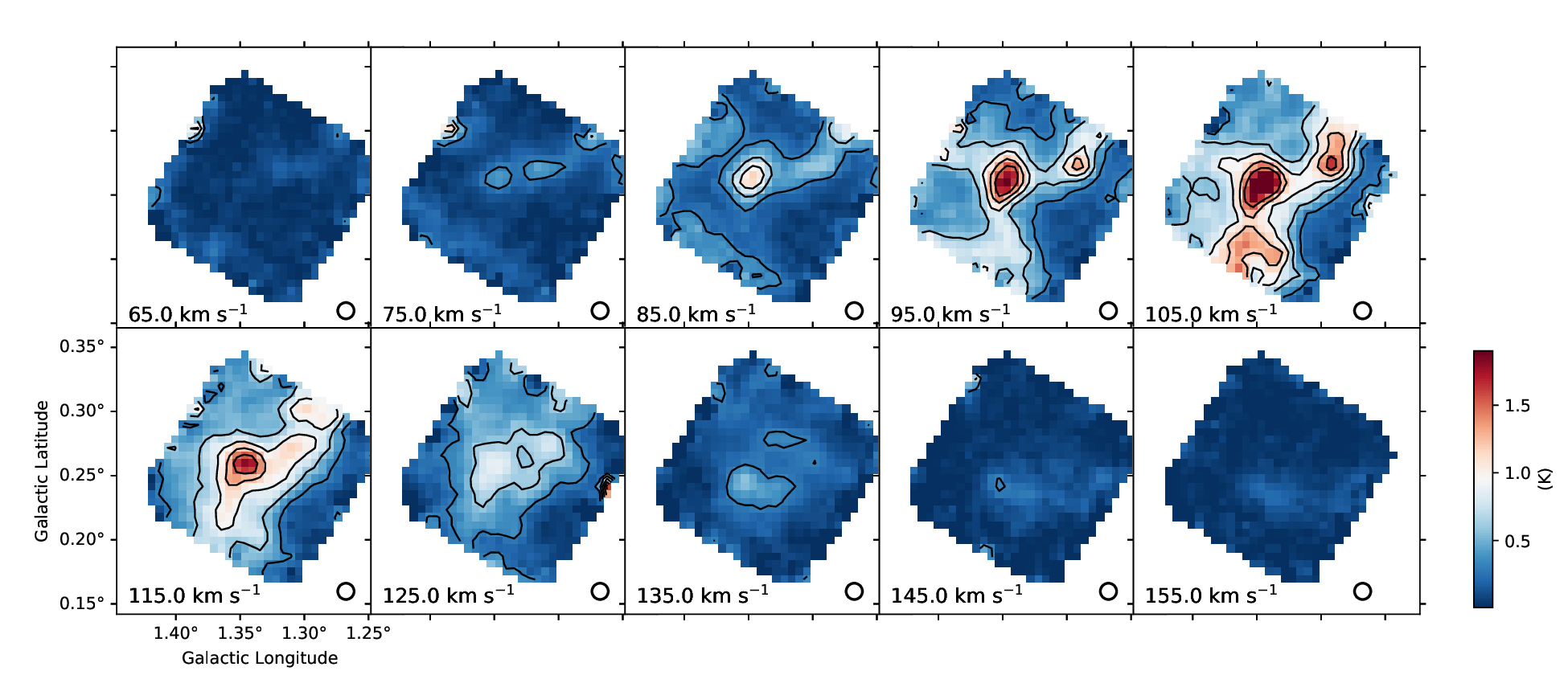}
\caption{Channel maps of SiO (1--0) emission at 43.4~GHz. Each panel shows the spatial distribution of emission at a velocity interval of 10 \kms, from 85 to 125 \kms. Contour levels are from 0.3~K to 1.5~K in steps of 0.3~K. The beam size is shown towards the bottom right of each panel.}
\label{FigA4}
\end{figure}

\begin{figure}[h]
\centering
\includegraphics[scale=0.4]{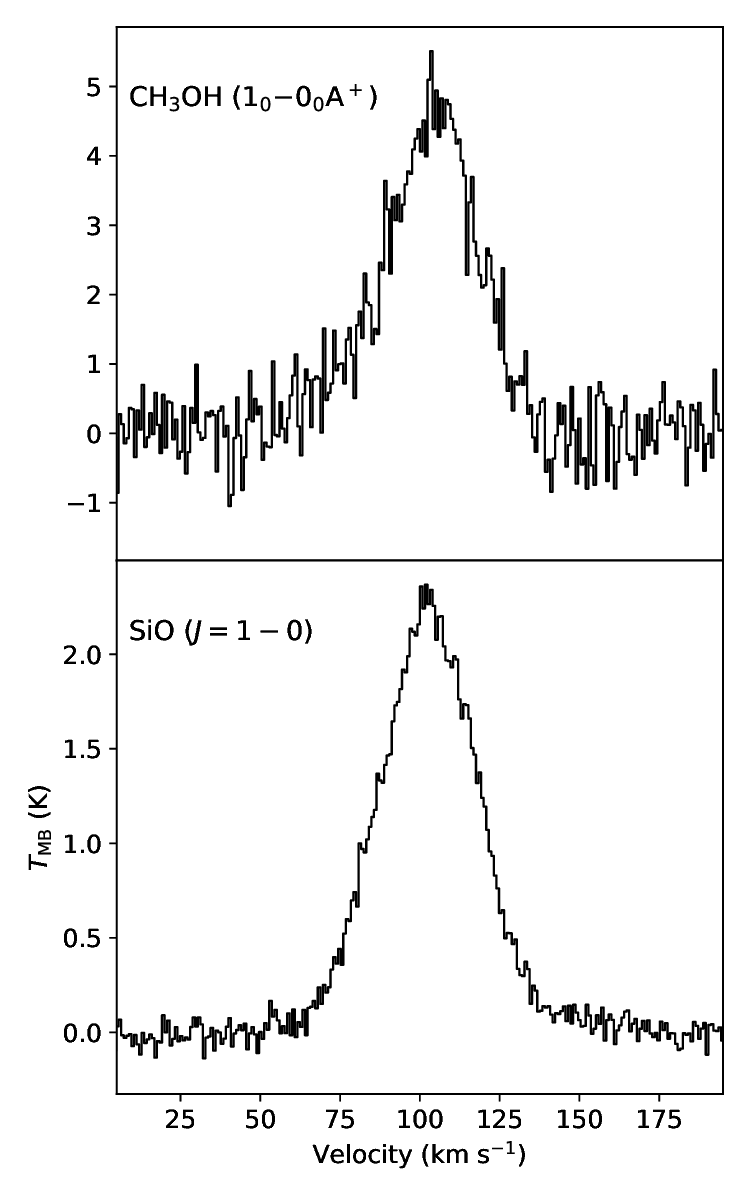}
\caption{Beam averaged spectra of CH$_3$OH emission at 48.4~GHz (top) and SiO emission at 43.4~GHz (bottom) towards the peak of the thermal-dominated emission at 36.2~GHz.}
\label{FigA5}
\end{figure}
\end{document}